\documentclass{article}
\usepackage{format}
\usepackage{latexsym}
\usepackage[english]{babel}
\usepackage{bm}
\usepackage[T1]{fontenc}
\usepackage[utf8]{inputenc}
\usepackage{csquotes}
\usepackage{ae}
\usepackage{graphicx}
\usepackage{subcaption}
\usepackage[graphicx]{realboxes}
\usepackage{epsfig}
\usepackage{color}
\usepackage{pgf}
\usepackage{pstricks,pst-grad}
\usepackage{float}
\usepackage{sidecap}
\sidecaptionvpos{figure}{c}
\usepackage[normalem]{ulem}
\usepackage{latexsym}
\usepackage{amsfonts}
\usepackage{amssymb}
\usepackage{amsthm}
\usepackage{amscd}
\usepackage{amsmath}
\usepackage{mathrsfs}
\usepackage{breqn}
\usepackage[bbgreekl]{mathbbol}
\usepackage{mathtools}
\usepackage{color}
\usepackage{colortbl}
\usepackage{framed}
\usepackage{tcolorbox}
\usepackage{verbatim}
\usepackage{pdfpages}
\usepackage{microtype}
\usepackage[inline]{enumitem}
\usepackage{verbatim}
\usepackage{placeins}
\usepackage[full]{textcomp}
\usepackage{wasysym}
\usepackage{appendix}
\usepackage[nolist,nohyperlinks]{acronym}
\usepackage{diagbox}
\usepackage{booktabs}
\usepackage{xr}
\usepackage{amsthm}
\usepackage{qcircuit}
\usepackage{physics}

\usepackage{hyperref}
\hypersetup{
    colorlinks=true,
    citecolor=black,
    linkcolor=red,
    linkbordercolor=blue,
}

\newcommand{\ie}{\it{i.e.}}
\newcommand{\g}{\text{glob}}
\newcommand{\gen}{\text{gen}}
\newcommand{\inst}{\text{inst}}

\begin{acronym}[MPTPS]
    \acro{QVAE}{quantum variational autoencoder}
    \acro{VAE}{variational autoencoder}
\end{acronym}
\usepackage[backend=biber,sorting=none]{biblatex}
\begin{document}

\renewcommand{\figureautorefname}{Fig.}
\renewcommand{\equationautorefname}{Eq.}
\renewcommand{\sectionautorefname}{Sec.}
\renewcommand{\subsectionautorefname}{subsec.}
\renewcommand{\tableautorefname}{Tab.}

\title{$\zeta$-QVAE: A Quantum Variational Autoencoder utilizing Regularized Mixed-state Latent Representations}

\author{%
Gaoyuan Wang$^{1,2 \dagger}$ \quad Jonathan Warrell $^{6,1,2 \dagger}$ \quad Prashant S. Emani$^{1,2}$ \quad Mark Gerstein$^{1,2,3,4,5*}$ \\ \\
$^1$ Program in Computational Biology and Bioinformatics,\quad \\
$^2$ Department of Molecular Biophysics and Biochemistry,\quad  \\
$^3$ Department of Computer Science,\quad  \\
$^4$ Department of Statistics \& Data Science,\quad  \\
$^5$ Department of Biomedical Informatics \& Data Science,\quad  \\
Yale University, New Haven, Connecticut 06520, USA \quad  \\
$^6$ NEC Laboratories America, Princeton, New Jersey 08540, USA \quad  \\
$\dagger \text{ These authors contributed equally to this work.}$\\
\texttt{*Corresponding author: mark@gersteinlab.org}\\
}
\maketitle

\begin{abstract}
A major challenge in near-term quantum computing is its application to large real-world datasets due to scarce quantum hardware resources.
One approach to enabling tractable quantum models for such datasets involves finding low-dimensional representations that preserve essential information for downstream analysis. 
In classical machine learning, variational autoencoders (VAEs) facilitate efficient data compression, representation learning for subsequent tasks, and novel data generation.
However, no quantum model has been proposed that exactly captures all of these features for direct application to quantum data on quantum computers.
Some existing quantum models for data compression lack regularization of latent representations, thus preventing direct use for generation and control of generalization.
Others are hybrid models with only some internal quantum components, impeding direct training on quantum data.
To address this, we present a fully quantum framework, $\zeta$-QVAE, which encompasses all the capabilities of classical VAEs and can be directly applied to map both classical and quantum data to a lower-dimensional space, while effectively reconstructing much of the original state from it. 
Our model utilizes regularized mixed states to attain optimal latent representations.
It accommodates various divergences for reconstruction and regularization. Furthermore, by accommodating mixed states at every stage, it can utilize the full-data density matrix and allow for a training objective defined on probabilistic mixtures of input data. Doing so, in turn, makes efficient optimization possible and has potential implications for private and federated learning.
In addition to exploring the theoretical properties of $\zeta$-QVAE, we demonstrate its performance on representative genomics and synthetic data.
Our results indicate that $\zeta$-QVAE consistently learns representations that better utilize the capacity of the latent space and exhibits similar or better performance compared to matched classical models.

\end{abstract}

\section{Introduction}

Autoencoders play an important role in current machine learning systems, enabling compression of data, learning latent representations, and, in certain cases, as generative models.  
Their architecture is comprised of an encoder that compresses the input into a lower dimensional intermediate latent state, and a decoder that reconstructs the input from the latent state.

Classical variational autoencoders (VAEs) provide a unified modeling framework which combines these strengths, and more recent classical models have extended this framework to allow a trade-off between reconstruction and information captured by the latent space \cite{higgins2016beta}, to maximize the coverage of the latent space and hence avoid generating spurious patterns \cite{tolstikhin2017wasserstein}, and to incorporate more complex encoders and decoders \cite{ronneberger2015u,ho2020denoising}

In the Noisy Intermediate-Scale Quantum (NISQ) era, quantum technologies are progressing rapidly, and classical machine learning methods are rapidly being generalized to operate in a quantum machine learning setting. Yet, the limited availability of quantum hardware and restrictions on the number of qubits in actual quantum devices underscores the need to minimize quantum resource requirements. In this work, we introduce a fully generalized \ac{QVAE} framework, which answers the challenges above by allowing efficient quantum data compression.  Our framework preserves or generalizes all the key features of classical VAE models, while directly operating on quantum data, to which classical compression methods cannot be directly applied. Notably, our proposed framework is valuable not just for quantum datasets but also for classical datasets due to the following potential advantages: (1) Quantum superposition offers the inherent advantage of a much richer representation space than classical binary bits. This enables potentially more efficient representations of data, crucial for compression into a compact latent space;
(2) The entanglement of qubits can be utilized to capture intricate dependencies in the original data via the encoding into latent states, which classical methods may be unable to represent efficiently; 
(3) Our framework employs quantum probability in place of classical distributions; for instance, we replace the classical Gaussian distributions typically used in VAEs with quantum mixed states.

A large number of proposals have been made to provide quantum analogues of autoencoder models  \cite{romero2017quantum,MA2023110659,Mangini2022,Bravo_Prieto_2021,Rivas2021}.  
Mostly, such models learn a quantum circuit to directly maximize the reconstruction of input quantum states. These pioneering models have demonstrated the ability to effectively compress quantum states. However, they come with several shortcomings. Such quantum autoencoder analogs are optimized for the reconstruction of quantum states but lack a regularization term over the latent space. Incorporating regularization with an isotropic prior can be advantageous, as it encourages the model to learn latent representations that utilize a larger portion of the latent space. Furthermore, without regularization, these models cannot be directly used for generation and do not provide explicit control over generalization.
We note that such models are `variational' in the sense that their quantum circuits may be trained using an approximate {\it Ansatz}, which differs from the approximation of a prior distribution which induces the regularization term in a classical VAE objective; we will thus refer to this type of model as a Quantum Autoencoder (QAE).
Further, the training for such QAEs assumes a particular form for the reconstruction error (quantum fidelity), and hence cannot be directly generalized if other forms of objective are required. In the quantum context, many different measures of similarity between quantum states have been proposed in addition to fidelity; the restriction to a particular similarity measure is thus undesirable.  Another shortcoming of such QAEs is that the input, encoded and output states are all assumed to be pure states; this sacrifices a unique potential advantage of quantum models for handling large datasets in parallel and embedding information using mixed quantum states.

An alternative kind of model is a hybrid quantum-classical analogue, such as the Ref.\,\cite{khoshaman2018quantum}, which is a classical VAE with a quantum Boltzmann distribution imposed upon its latent variables. Although such hybrid models are trained as generative models, their objective is defined using a bound on the classical log-likelihood (hence they are trained to generate and reconstruct classical data).  They cannot therefore be trained directly on quantum inputs, and moreover sacrifice the potential virtues of handling data efficiently through mixed quantum states. Hence, none of the current models is able to integrate the advantages of mixed-state input and latent representations, flexible objectives and effective, fully quantum regularization. 
However, such capabilities are particularly important for scaling up quantum models to handle real-world datasets with large feature spaces by compressing them down to dimensions feasible for NISQ quantum hardware.

We therefore introduce a gate-based quantum variational autoencoder framework with a training objective that includes a latent space regularization term (and hence may be viewed as a generalized form of probabilistic generative models). We propose using the maximally mixed state as an isotropic prior on the latent state for regularization. 
This approach enhances the utilization of the latent space's representation capacity, improves the preservation of the relational structure of the data points, and increases their applicability for downstream analysis.
Additionally, we introduced a regularization coefficient to control the influence of the prior.
We refer to our model as a $\zeta$-QVAE, where $\zeta$ represents the density matrix of the mixed-state latent representation in our model (analogous to the classical latent state, $Z$). The $\zeta-$QVAE allows for regularized latent representations to exist as mixed states within the Bloch sphere, which further enriches the representation space, and potentially allows more efficient data compression.
In our framework, the encoder and decoder pairs are modeled as quantum operations (completely positive trace-preserving (CPTP) linear maps), to provide mixed-state latent representations for quantum inputs (which may be mixed or pure states). 
Our model also offers multiple divergence options for the reconstruction and regularization losses, since specific losses are commonly used for particular applications of Quantum Machine Learning.
Specifically, we show how i) fidelity \cite{doi:10.1080/09500349414552171}, frequently utilized in quantum state tomography \cite{Huang_2020}, ii) quantum relative entropy/quantum Jensen-Shannon divergence, significant in the field of quantum information theory \cite{RevModPhys.74.197,fawzi2018efficient,Majtey_2005}, and iii) quantum Wasserstein distance, applied in generative models \cite{chakrabarti2019quantum}, can be integrated into our framework.
Furthermore, we show that a quantum information-theoretic analogue to the classical evidence lower bound (ELBO) exists for the regularized reconstruction loss when the quantum relative entropy is used.

Moreover, we formulate both local and global versions of each divergence, which allow models to be optimized for reconstruction of individual data points or probabilistic mixtures of data points, respectively. In particular, the quantum Wasserstein distance can be shown to give rise to equivalent optimal models under both global and instance-based objectives, while the other global divergences were observed to give similarly good performance under both on a real-world genomics dataset. The global version of each divergence allows efficient optimization by reducing the number of repeated quantum operations during the training. It also has implications for private and federated learning, since it requires only aggregate information about the dataset as opposed to individual data points. Further, we discuss suitable application domains for the $\zeta$-QVAE, as well as addressing the challenges associated with implementing our framework on NISQ hardware.
We wish to note that the methods discussed in this work presume the availability of algorithms for the efficient encoding of input states, as well as the storage of the latent states. While an exploration of preparation and storage algorithms is beyond the scope of this manuscript, we discuss them briefly in Section 7 while considering implementation in NISQ devices.

\section{Theoretical Framework}\label{sec_2}

We assume that our data live in an input Hilbert space, $X$, over $N_X$ qubits, and that we wish to learn an encoder/decoder pair to compress our dataset to Hilbert space, $Z$, over $N_Z$ qubits ($N_Z \leq N_X$).  Our dataset consists of a finite set of $N$ pure states, $\ket{\psi_1}...\ket{\psi_N}$ (which may themselves be generated from classical data-points, e.g. by amplitude encoding, or may be generated from an intrinsically quantum source).  We use the following definitions for the input density matrices of individual datapoints (indexed by $i$), and a global density matrix representing the entire dataset:

\begin{eqnarray}
\rho_i &=& \ket{\psi_i}\bra{\psi_i} \nonumber \\
\rho_\g &=& \frac{1}{N} \sum_i \rho_i
\label{eq:def_global_state}
 \end{eqnarray}

\noindent We note that our dataset may be considered a finite sample from a distribution across pure states over $X$, and hence $\rho_\g$ may be considered a finite-sample approximation to an underlying data distribution.  Our goal is to learn quantum operations (completely positive trace-preserving (CPTP) linear maps) $(\mathcal{E},\mathcal{D})$ corresponding to an encoder and decoder respectively, where these have the respective signatures $\mathcal{E}:D(X)\rightarrow D(Z)$ and $\mathcal{D}:D(Z)\rightarrow D(X)$ (here, $D(X)$ denotes the set of density matrices over finite Hilbert space $X$; we note also that, due to circuit constraints, we may have $\mathcal{E}\in S_{\mathcal{E}}$ and $\mathcal{D}\in S_{\mathcal{D}}$, where $S_{\mathcal{E}}$ and $S_{\mathcal{D}}$ are subsets of CPTP linear maps having a predefined maximum circuit complexity).  Given an $(\mathcal{E},\mathcal{D})$ pair, we define:

\begin{eqnarray}\label{eq_encDecDef1}
\zeta_i &=& \mathcal{E}(\rho_i) \nonumber \\
\sigma_i &=& \mathcal{D}(\zeta_i)
 \end{eqnarray}

\noindent where $\zeta_i$ and $\sigma_i$ represent the latent and reconstructed state respectively associated with input state $\rho_i$, and similarly, $\zeta_\g = \mathcal{E}(\rho_\g)$ and $\sigma_\g = \mathcal{D}(\zeta_\g)$.  Further, we assume we have a predefined `prior' density matrix over the latent space, $\zeta_\gen$; below, we will take this to be the maximally mixed state, $\zeta_\gen=(1/2^{N_Z})I_{2^{N_Z}}$. As in the classical case, this prior, $\zeta_\gen$, is transformed by the decoder to produce a generative approximation of the data distribution, $\sigma_\gen = \mathcal{D}(\zeta_\gen)$.  For concreteness, to define our encoder and decoder (see \autoref{fig:Full} and \autoref{fig:Encoder_circuit}), we append $N_A$ auxiliary qubits to our input Hilbert space $X$, and $N_T$ reference qubits along with $N_B$ auxiliary qubits to our latent Hilbert space $Z$ ($N_T = N_X - N_Z$, where $N_T$ denotes the number of `trash' qubits).  Then, we can use the following definitions:

\begin{eqnarray}\label{eq_encDecDef}
\mathcal{E}(\rho) &=& \Tr_{N_A+N_T}(U^{-1}(\rho\otimes \ket{0_{N_A}}\bra{0_{N_A}})U) \nonumber \\
\mathcal{D}(\zeta) &=& \Tr_{N_B}(V^{-1}(\zeta\otimes \ket{0_{N_B+N_T}}\bra{0_{N_B+N_T}})V) 
 \end{eqnarray}

\noindent where $U$ and $V$ are unitary matrix representations of the encoder and decoder circuits ($\mathcal{E}$ and $\mathcal{D}$) respectively, and $\Tr_N(.)$ denotes the trace over the final $N$ qubits (where, in general, the qubits may be ordered/indexed arbitrarily, although below we assume that the auxiliary qubits are ordered after those in $X$ and $Z$).  In Appendix A (Prop. 1), we prove that setting $N_A = N_B = N_X + 2N_Z$ is sufficient to allow arbitrary pairs of quantum operations $(\mathcal{E},\mathcal{D})$ to be learned (assuming no circuit complexity constraints).

\vspace{3mm}
\noindent \textbf{Global Training Objective:} To derive a training loss for the model above, we assume that we are interested in learning a model which minimizes a completely general loss $\mathcal{L}_1(a,b)$ (the only assumptions being that it is non-negative and 0 iff $a=b$, but not necessarily symmetric, $\ie,$ a divergence) between the implicit generative model and the global data density matrix (note that we derive an alternative instance-based objective below); hence we seek to optimize:

\begin{eqnarray}\label{eq_glob}
\min_{\mathcal{D}} \mathcal{L}_1(\rho_\g,\sigma_\gen)
 \end{eqnarray}

\noindent We note that \autoref{eq_glob} involves only the decoder, $\mathcal{D}$.  In analogy with the classical VAE, to simultaneously learn a representation of our data in the latent space, we introduce a variational density parameterized  by our encoder $\mathcal{E}$, which we assume (temporarily) to be expressive enough to fulfill the condition $\zeta_\g=\zeta_\gen$:

\begin{eqnarray}\label{eq_var}
\min_{\mathcal{D}} \mathcal{L}_1(\rho_\g,\sigma_\gen) = \min_{\substack{\mathcal{E},\mathcal{D} \\ \text{s.t.} \zeta_\g=\zeta_\gen}} \mathcal{L}_1(\rho_\g,\sigma_\g)
 \end{eqnarray} 

\noindent We can reformulate \autoref{eq_var} as a constrained optimization problem, introducing a second (regularization) loss $\mathcal{L}_2$ (with the same conditions as $\mathcal{L}_1$):

\begin{eqnarray}\label{eq_constr}
\min_{\mathcal{E},\mathcal{D}} \mathcal{L}_1(\rho_\g,\sigma_\g) \nonumber \\
\mathcal{L}_2(\zeta_\g,\zeta_\gen) \leq \epsilon
 \end{eqnarray} 

\noindent To account for the fact that our class of encoders may not allow $\mathcal{L}_2(\zeta_\g,\zeta_\gen)=0$ to be fulfilled, we introduce the constant $\epsilon=\min_{\mathcal{E}}(\mathcal{L}_2(\zeta_\g,\zeta_\gen))$ in \autoref{eq_constr}.  Finally, introducing the Lagrange multiplier $\beta\geq 0$, we derive our training objective $F$ for a global input density matrix, $\rho_\g$:

\begin{eqnarray}\label{eq_lagr}
\min_{\substack{\mathcal{E},\mathcal{D} \\ \text{s.t.} \mathcal{L}_2(\zeta_\g,\zeta_\gen)\leq \epsilon}} \mathcal{L}_1(\rho_\g,\sigma_\g) &\geq& \max_\beta \min_{\mathcal{E},\mathcal{D}} F_\g(\mathcal{E},\mathcal{D},\beta) \nonumber \\
F_\g(\mathcal{E},\mathcal{D},\beta) &=& \mathcal{L}_1(\rho_\g,\sigma_\g) + \beta (\mathcal{L}_2(\zeta_\g,\zeta_\gen)-\epsilon)
 \end{eqnarray} 

\vspace{3mm}
\noindent In place of the maximization across $\beta$ on the RHS (upper) of \autoref{eq_lagr}, we treat $\beta$ as a hyperparameter when optimizing $F_\g$, and we disregard the constant $-\beta\epsilon$ for the objectives used in the $\zeta$-QVAE (as defined in \autoref{eq_objs} and \autoref{eq_objsAlt}).  We also show, in Appendix A (Prop. 2), that when $\mathcal{L}_1$ and $\mathcal{L}_2$ are the quantum relative entropy, $\beta=1$ and $\epsilon=0$, that $F_\g(\mathcal{E},\mathcal{D},\beta)$ forms an analogue of the classical Evidence Lower-Bound (ELBO), as in the classical VAE \cite{kingma2013auto} (we note that this bound is distinct from the Q-ELBO bound in \cite{khoshaman2018quantum}, since the Q-ELBO is a bound on the classical log-likelihood, while Prop. 2 is a bound on the quantum relative entropy).  The $\zeta$-QVAE objective therefore optimizes the original objective in \autoref{eq_glob} in the case that either $\mathcal{L}_1$ and $\mathcal{L}_2$ are the quantum relative entropy with $\beta=1$, or $\beta=\beta^*$, where $\beta^*$ is the optimum value of $\beta$ in the RHS (upper) of \autoref{eq_lagr}.

\vspace{3mm}
\noindent \textbf{Instance-based Training Objective:} In the above, \autoref{eq_lagr} provides an objective for training $(\mathcal{E},\mathcal{D})$ based on the generation and reconstruction of the global data density matrix $\rho_\g$.  
However, we are also interested in the setting where the reconstruction of individual data points is directly optimized; this is not explicitly represented in \autoref{eq_lagr}, in contrast to conventional classical VAE and previous QAE frameworks.
For this reason, we consider the following optimization problem for instance-level generation/reconstruction:

\begin{eqnarray}\label{eq_inst}
\min_{\mathcal{D}} \sum_i \mathcal{L}_1(\rho_i,\sigma_\gen)
 \end{eqnarray} 

\noindent By a similar argument to above, this leads to the following instance-level objective, $F_\inst$:

\begin{eqnarray}\label{eq_lagrInst}
\min_{\substack{\mathcal{E},\mathcal{D} \\ \text{s.t.} \mathcal{L}_2(\zeta_i,\zeta_\gen)\leq \epsilon, \forall i}} \sum_i \mathcal{L}_1(\rho_i,\sigma_i) &\geq & \max_{\beta_{1...N}} \min_{\mathcal{E},\mathcal{D}} F_\inst(\mathcal{E},\mathcal{D},\beta_{1...N}) \nonumber \\
F_\inst(\mathcal{E},\mathcal{D},\beta_{1...N}) &=& \sum_i \left( \mathcal{L}_1(\rho_i,\sigma_i) + \beta_i (\mathcal{L}_2(\zeta_i,\zeta_\gen)-\epsilon) \right)
 \end{eqnarray} 

\noindent As above, we treat the $\beta$'s as a hyperparameter, using a common $\beta=\beta_1=...=\beta_N$, and ignore the constant terms $-\beta_i \epsilon$.  Here, $\epsilon=\min_{\mathcal{E}}(\max_i \mathcal{L}_2(\zeta_i,\zeta_\gen))$, and \autoref{eq_lagrInst} forms a strict lower-bound on \autoref{eq_inst} when $\epsilon=0$.  In general, $F_\g$ and $F_\inst$ will lead to different optimization problems, and hence different solutions for $(\mathcal{E},\mathcal{D})$; however, in Appendix A (Prop. 3), we show that for certain losses, the optimization problems in \autoref{eq_lagr} and \autoref{eq_lagrInst} become equivalent.

We note finally that, if we have an auxiliary loss function $\mathcal{L}'_1$ for which:

\begin{eqnarray}\label{eq_aux}
\min_{\substack{\mathcal{E},\mathcal{D} \\ \text{s.t.} \mathcal{D}(\mathcal{E}(\rho))=\sigma}} \mathcal{L}'_1(\rho,\mathcal{E},\mathcal{D}) = \mathcal{L}_1(\rho,\sigma),
 \end{eqnarray} 

\noindent we may use the following alternative definitions of $F_\g$ and $F_\inst$ in \autoref{eq_lagr} and \autoref{eq_lagrInst}:

\begin{eqnarray}\label{eq_lagrAlt}
F'_\g(\mathcal{E},\mathcal{D},\beta) &=& \mathcal{L}'_1(\rho_\g,\mathcal{E},\mathcal{D}) + \beta (\mathcal{L}_2(\zeta_\g,\zeta_\gen)-\epsilon) \nonumber \\
F'_\inst(\mathcal{E},\mathcal{D},\beta_{1...N}) &=& \sum_i \left( \mathcal{L}'_1(\rho_i,\mathcal{E},\mathcal{D}) + \beta_i (\mathcal{L}_2(\zeta_i,\zeta_\gen)-\epsilon) \right)
 \end{eqnarray} 

 \noindent This version of the bound will used below when $\mathcal{L}_1$ is the Wasserstein divergence.

\section{Model}
\subsection{$\zeta$-QVAE architecture}

\begin{figure}[h]
\centering
\begin{subfigure}{0.5\textwidth}
\includegraphics[width=1\linewidth]{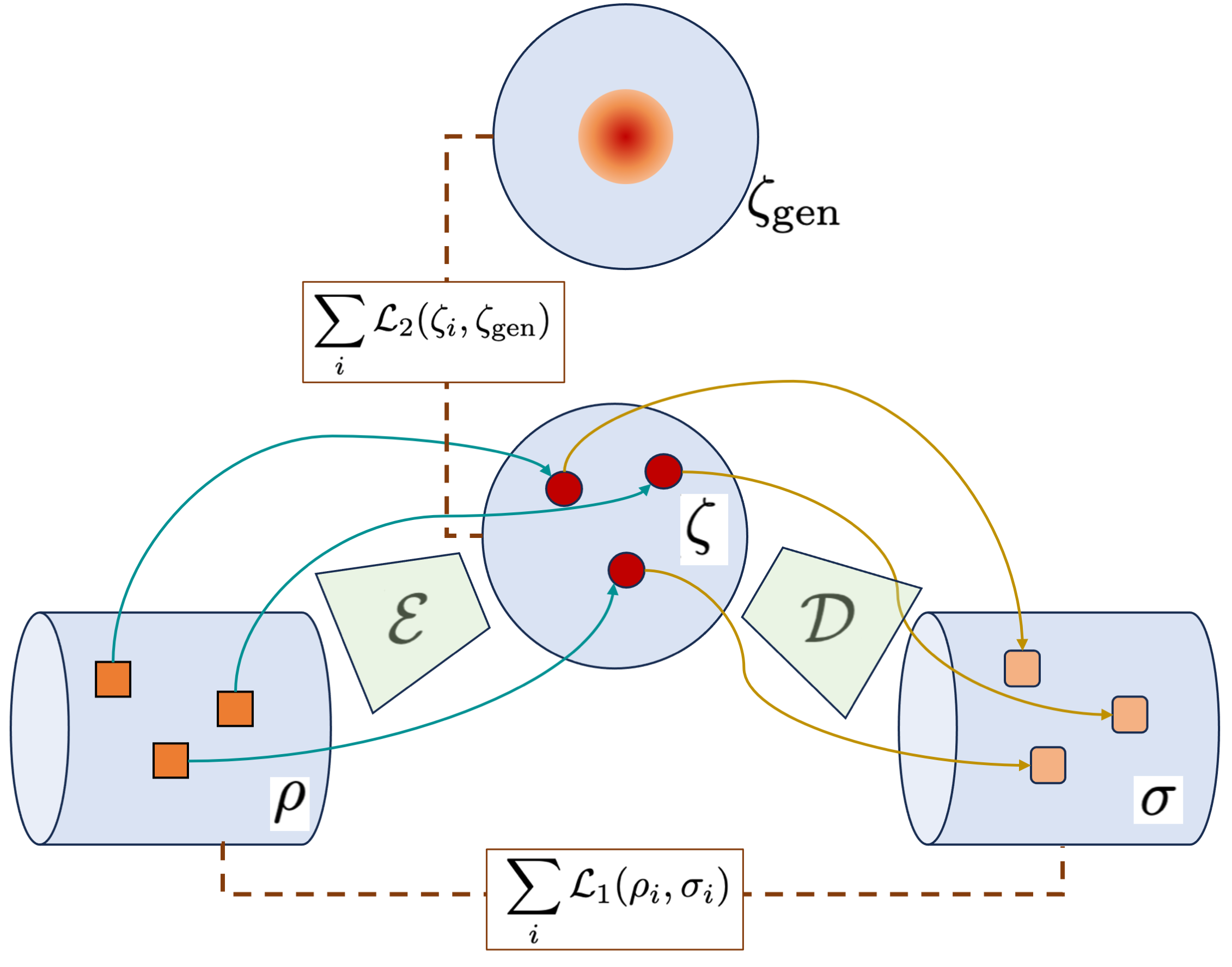}
\caption{}
\end{subfigure}
\begin{subfigure}{0.85\textwidth}
\includegraphics[width=1\linewidth]{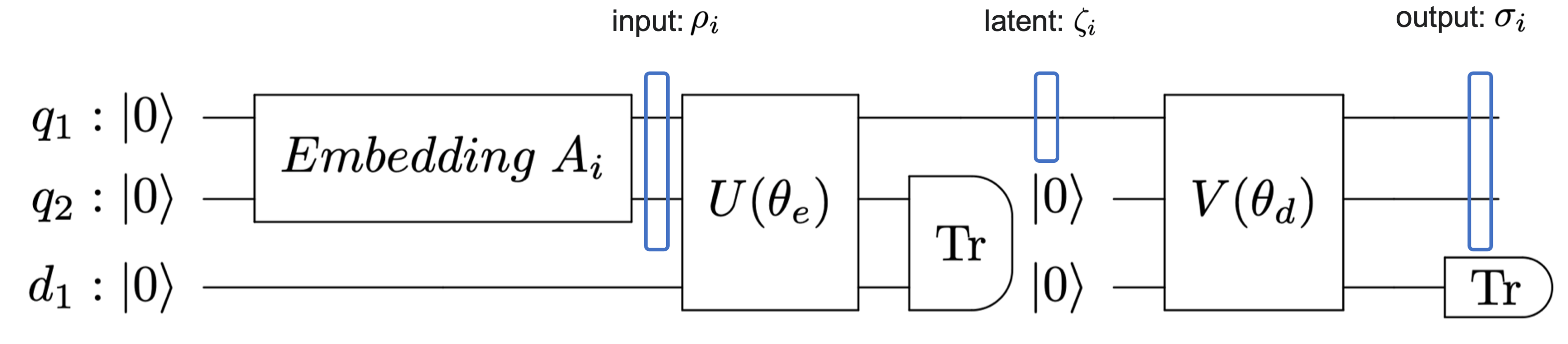}
\caption{}
\end{subfigure}
\caption{(a) Illustration of the architecture and objective function of the $\zeta$-QVAE; arrows represent transformations between mixed states, $\zeta_\gen$ is the maximally mixed state, and  $\mathcal{L}_1$ and $\mathcal{L}_2$ are the reconstruction and regularization loss respectively. (b) $\zeta$-QVAE overall circuit representation.}
\label{fig:Full}
\end{figure}

The overall architecture of the proposed \ac{QVAE} is given in \autoref{fig:Full} with an example of $N_X=2$ input qubits, a latent space of $N_Z=1$ qubit, and one auxiliary qubit ($d_1$) in both the encoder and decoder (hence, $N_A=N_B=1$). The encoder and decoder are defined by quantum circuits, with trainable parameters $\theta_e$ and $\theta_d$ respectively.  The corresponding unitary matrices are denoted $U(\theta_e)$ and $V(\theta_d)$ respectively.  Additionally, the unitary matrix $A_i$ performs the conversion from a classical source to a quantum representation for data-point $i$ (e.g. using amplitude or angle embedding).  Hence, $\ket{\psi_i} = A_i \ket{0}$, and $\rho_i=\ket{\psi_i}\bra{\psi_i}$.  After the embedding and encoder circuits have been applied to the initial $\ket{0}$ state, both the $N_{A}$ auxiliary qubits and the $N_T=N_X-N_Z$ trash qubits are discarded by a partial trace operation, and the remaining qubit $q_1$ is considered the latent state. The encoder $\mathcal{E}$
as a whole therefore has the form defined by \autoref{eq_encDecDef}.  To reconstruct original information from the latent state,  $N_T + N_{B}$ zero state qubits are added to the remaining qubits, and a final partial trace is performed across the auxiliary qubit $d_1$; hence the decoder $\mathcal{D}$ is of the form in \autoref{eq_encDecDef}.

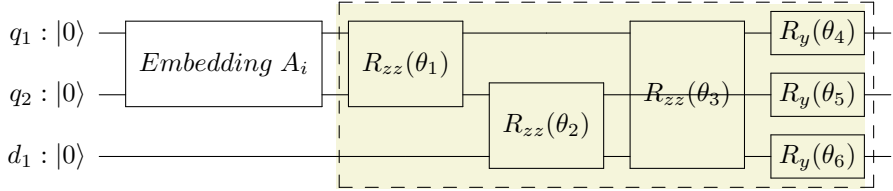
\begin{figure}[h]
\centering
\scalebox{1}{
\begin{pgfpicture}{0em}{0em}{0em}{0em}
\definecolor{beige}{rgb}{0.96, 0.96, 0.86}
\color{beige}
\pgfrect[fill]{\pgfpoint{10.3em}{-5.9em}}{\pgfpoint{20em}{7em}}
\end{pgfpicture}
\Qcircuit @C=1em @R=.7em {
& \lstick{q_1: \ket{0}}  &\multigate{1}{Embedding \; A_i}& \multigate{1}{R_{zz}(\theta_1)}&\qw& \multigate{2}{R_{zz}(\theta_3)}&\gate{R_y(\theta_4)}&\qw\\
& \lstick{q_2: \ket{0}} &\ghost{Embedding \; A_i} & \ghost{R_{zz}(\theta_1)} &\multigate{1}{R_{zz}(\theta_2)}&\qw&\gate{R_y(\theta_5)}&\qw\\
& \lstick{d_1: \ket{0}}  &\qw & \qw &  \ghost{R_{zz}(\theta_2)}&  \ghost{R_{zz}(\theta_3)}&\gate{R_y(\theta_6)}&\qw
\gategroup{1}{4}{3}{7}{.7em}{--}
}
}
\caption{Encoder circuit}
\label{fig:Encoder_circuit}
\end{figure}

In \autoref{fig:Encoder_circuit}, the encoder circuit $U(\theta_e)$ we used in this study is shown for one trainable layer $\ie\,N_l=1$ (note that we use a data embedding circuit, $A_i$, to first project classical data to a quantum state, e.g. via amplitude encoding; this is not formally part of the $\zeta$-QVAE encoder, and can be removed if a quantum data source provides the input state). The Ansatz (marked in beige) was introduced by \cite{lloyd2020quantum} and contains $R_{zz}$ entangling gates and single qubit $R_y$ rotations. The decoder circuit contains the same Ansatz as the encoder.

\vspace{1em}

\subsection{Training objectives}
\label{sec:training_objectives}

\textbf{Reconstruction loss, $\mathcal{L}_1$}:
We provide here the explicit forms of all the divergences we consider for the reconstruction loss.  As in \autoref{sec_2}, a divergence between two density matrices $\rho$ and $\sigma$ over the same Hilbert space is a non-negative function, $\mathcal{L}(\rho,\sigma)$, which is zero iff $\rho=\sigma$, but unlike a metric, need not be symmetric.  For generality, we write all divergences below for arbitrary $\rho$ and $\sigma$.  However, we are particularly interested in the cases $\mathcal{L}(\rho_\g,\sigma_\g)$ and $\mathcal{L}(\rho_i,\sigma_i)$, denoting the divergence between input and output density matrices for the global and instance level objectives respectively (see \autoref{eq_lagr} and \autoref{eq_lagrInst}).  For these cases, $\rho_i= \ket{\psi_i}\bra{\psi_i}$, where $\ket{\psi_i}$ is the state vector of the $i$-th input data-point, $\rho_\g = (1/N)\sum_i \rho_i$, $\sigma_i=\mathcal{D}(\mathcal{E}(\rho_i))$, and $\sigma_\g=\mathcal{D}(\mathcal{E}(\rho_\g))$, where $\mathcal{E}$ and $\mathcal{D}$ are the quantum operation representations of the encoder and decoder, as in \autoref{sec_2}.  The particular losses we consider for the reconstruction loss, $\mathcal{L}_1(\rho,\sigma)$, are summarized below:

\vspace{3mm}
\begin{itemize}
    \item Fidelity loss:
    \begin{eqnarray}
        \mathcal{L}^{f}_1(\rho,\sigma) = 1 - \left( \Tr \sqrt{\sqrt{\sigma} \rho \sqrt{\sigma}} \right)^2,
    \end{eqnarray}
    where, for a pure state $\rho=\ket{\psi_\rho}\bra{\psi_\rho}$, this reduces to: $\mathcal{L}^{f}_1(\rho,\sigma) = \bra{\psi_\rho}\sigma\ket{\psi_\rho}$.
    
    \item Quantum relative entropy (KLD):
    \begin{eqnarray}
        \mathcal{L}^{kl}_1(\rho,\sigma) = S(\rho|\sigma) = S(\rho,\sigma) - S(\rho) = - \Tr (\rho\log(\sigma)) - S(\rho),
    \end{eqnarray}
    where $S(\rho)=-\Tr (\rho\log(\rho))$ and $S(\rho,\sigma)=-\Tr (\rho\log(\sigma))$ .

    \item Symmetric quantum relative entropy (JSD)  \cite{Majtey_2005}:
        \begin{eqnarray}
    \mathcal{L}^{jsd}_1(\rho,\sigma)   &=& S\Big(\rho \;\Big|\; \frac{1}{2}\left[\rho+\sigma\right]\Big)+S\Big(\sigma \;\Big|\; \frac{1}{2}\left[\rho+\sigma\right]\Big).
    \end{eqnarray}
   
    \item Quantum Wasserstein-distance loss: 
        \begin{eqnarray}
            \mathcal{L}^{w}_1(\rho,\sigma)= \min_{\mathcal{T}:\mathcal{T}(\rho)=\sigma} \Tr(\pi(\rho,\mathcal{T}) C),
    \end{eqnarray}
    where $\mathcal{T}$ is a quantum operation, and:
    \begin{eqnarray}
        \pi(\rho,\mathcal{T}) &\coloneqq& \sum_i p_i (\mathcal{T}(\ket{e_i}\bra{e_i}))\otimes(\ket{e_i}\bra{e_i})
    \end{eqnarray}
    with $\ket{e_i}$ an orthogonal basis for $\rho$, hence $\rho =\sum_i p_i\ket{e_i}\bra{e_i}$, and $C$ is defined as in \cite{chakrabarti2019quantum}.  As discussed in \autoref{sec_2}, we introduce the following auxiliary loss function in place of $\mathcal{L}^{w}_1$ for the reconstruction loss when using the Quantum Wasserstein-distance:
        \begin{eqnarray}\label{eq_altQW}
            (\mathcal{L}'_1)^{w}(\rho,\mathcal{E},\mathcal{D})= \Tr(\pi(\rho,\mathcal{D}\circ\mathcal{E}) C),
    \end{eqnarray}
    Clearly, we have:
    \begin{eqnarray}
            \min_{\substack{\mathcal{E},\mathcal{D} \\ \text{s.t.} \mathcal{D}(\mathcal{E}(\rho))=\sigma}} (\mathcal{L}'_1)^{w}(\rho,\mathcal{E},\mathcal{D}) = \mathcal{L}^{w}_1(\rho,\sigma),
    \end{eqnarray}
    and so we can use the alternative form of the training objectives in \autoref{eq_lagrAlt} to optimize $\mathcal{L}^{w}_1(\rho,\sigma)$.

\end{itemize}

\vspace{3mm}
\noindent \textbf{Regularization loss, $\mathcal{L}_2$}: We write the regularization loss below in the general form $\mathcal{L}_2(\zeta,\zeta_\gen)$, i.e. a divergence between a mixed-state latent representation $\zeta$ and the analog of the classical generative `prior' on the latent space, $\zeta_\gen$.  As discussed in \autoref{sec_2}, we use $\zeta_\gen=\frac{1}{\lambda} I$, where $I$ is the identity operator and $\lambda$ the dimension of the latent Hilbert space.  This represents the maximally mixed state, $\ie\,$ the quantum state with the maximal entropy. In principle, all the divergences above could be used for the regularization loss, $\mathcal{L}_2$.  However, we exclude the quantum Wasserstein loss, since this would require us to minimize over an auxiliary circuit to find the lowest-cost transformation $\mathcal{T}$ between $\zeta$ and $\zeta_\gen$.  We briefly summarize the remaining divergences used for $\mathcal{L}_2$, with the simplifications induced by setting $\zeta_\gen=\frac{1}{\lambda} I$.
 
\begin{itemize}

    \item Fidelity loss:
    \begin{eqnarray}
        \mathcal{L}^{f}_2(\zeta,\zeta_\gen) = 1 - \left( \Tr \sqrt{\sqrt{\zeta_\gen} \zeta \sqrt{\zeta_\gen}} \right)^2.
    \end{eqnarray}
    
    \item Quantum relative entropy (KLD): 
    \begin{eqnarray}
        \mathcal{L}^{kl}_2(\zeta,\zeta_\gen) = S(\zeta,\zeta_\gen)-S(\zeta) = \Tr (\zeta\log(\zeta)) - \log(1/\lambda) = -S(\zeta)+c,
    \end{eqnarray}
    where $c=- \log(1/\lambda)$.

    \item Symmetric quantum relative entropy (JSD):
    \begin{eqnarray}
    \mathcal{L}^{jsd}_2(\zeta,\zeta_\gen)   &=& S\Big(\zeta \;\Big|\; \frac{1}{2}\left[\zeta+\zeta_\gen\right]\Big)+S\Big(\zeta_\gen \;\Big|\; \frac{1}{2}\left[\zeta+\zeta_\gen\right]\Big)
    \end{eqnarray}

\end{itemize}

\vspace{3mm}
\noindent \textbf{Overall training objectives}:  For explicitness, we collect together the specific forms of the overall global and instance based training objectives used to train our model, based on \autoref{eq_lagr} and \autoref{eq_lagrInst} respectively:

\begin{eqnarray}\label{eq_objs}
\mathcal{L}_\g(\theta_e,\theta_d,\beta) &=& \mathcal{L}_1(\rho_\g,\sigma_\g) + \beta \mathcal{L}_2(\zeta_\g,\zeta_\gen) \nonumber \\
\mathcal{L}_\inst(\theta_e,\theta_d,\beta) &=& \sum_i \left( \mathcal{L}_1(\rho_i,\sigma_i) + \beta \mathcal{L}_2(\zeta_i,\zeta_\gen) \right)
 \end{eqnarray} 
 
\noindent along with the alternative forms used for the Wasserstein reconstruction loss based on \autoref{eq_lagrAlt} and \autoref{eq_altQW}:

\begin{eqnarray}\label{eq_objsAlt}
\mathcal{L}'_\g(\theta_e,\theta_d,\beta) &=& \mathcal{L}'_1(\rho_\g,\mathcal{E}(\theta_e),\mathcal{D}(\theta_d)) + \beta \mathcal{L}_2(\zeta_\g,\zeta_\gen) \nonumber \\
\mathcal{L}'_\inst(\theta_e,\theta_d,\beta) &=& \sum_i \left( \mathcal{L}'_1(\rho_i,\mathcal{E}(\theta_e),\mathcal{D}(\theta_d)) + \beta \mathcal{L}_2(\zeta_i,\zeta_\gen) \right).
 \end{eqnarray} 
We are motivated to introduce regularization into the objective function in order to improve the quality and applicability of the learned latent representations. Particularly, by tuning the parameter $\beta$, we aim to achieve the optimal balance between the reconstruction fidelity and quality of latent representations. In this work, we use the Pearson correlation coefficient between the distance matrices of the input and latent states, and performance on downstream classification tasks using the latent states, as metrics for assessing the quality and applicability of the latent representations learned by our $\zeta$-QVAE. 

\subsection{QSVC classifier}

In addition to the ability of the $\zeta$-QVAE to reconstruct the original states, we are also interested in how well the latent and reconstructed states belonging to different classes can be effectively distinguished. In other words, we want to evaluate the classification performance on the latent and reconstructed states in comparison to the original input states. 
To evaluate this, we implemented a quantum kernel based classifier\cite{lloyd2020quantum} with amplitude embedding. The classifier is used to perform downstream analysis and is trained independently of the $\zeta$-QVAE. It is applied to the output of the $\zeta$-QVAE after the $\zeta$-QVAE training is complete.

\begin{figure}[h]
\centering
\scalebox{0.8}{
\centering
\begin{pgfpicture}{0em}{0em}{0em}{0em}
\definecolor{beige}{rgb}{0.96, 0.96, 0.86}
\color{beige}
\pgfrect[fill]{\pgfpoint{10.1em}{-5.9em}}{\pgfpoint{20.3em}{7em}}
\end{pgfpicture}
\Qcircuit @C=1em @R=.7em {
& \lstick{q_1: \ket{0}} & \multigate{2}{Embedding \; A_i} & \multigate{1}{R_{zz}(\theta_1)}&\qw& \multigate{2}{R_{zz}(\theta_3)}&\gate{R_y(\theta_4)} &  \multigate{2}{QSVC}&\qw\\
& \lstick{q_2: \ket{0}} &   \ghost{Embedding \; A_i} & \ghost{R_{zz}(\theta_1)} &\multigate{1}{R_{zz}(\theta_2)}&\qw&\gate{R_y(\theta_5)}&\ghost{QSVC}&\qw\\
& \lstick{q_3: \ket{0}} &  \ghost{Embedding \; A_i} & \qw &  \ghost{R_{zz}(\theta_2)}&  \ghost{R_{zz}(\theta_3)}&\gate{R_y(\theta_6)}&\ghost{QSVC}&\qw
\gategroup{1}{4}{3}{7}{.7em}{--}
}
}
\caption{The overall QSVC architecture.}
\label{fig:classifier_circuit}
\end{figure}
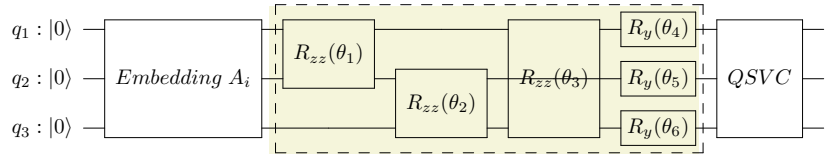

Our QSVC classifier is illustrated in \autoref{fig:classifier_circuit} using an example with one trainable layer. We use the same Ansatz, which includes alternating $R_{zz}$ and $R_y$ gates for the quantum kernel, as employed in the encoder and decoder.
The similarity kernel of our QSVC is obtained as the quantum fidelity between each data pair.
Due to the nature of the gene expression data and the normalizations we applied to our data for the amplitude embedding, we have observed a concentration of fidelity scores towards the higher end rather than being spread across the entire range from zero to one. To address this, we introduced a scaling function $$f\Big(\bra{v_i}\ket{v_j}\Big)=\tan\Big(\frac{\pi}{2.03}\bra{v_i}\ket{v_j}\Big)$$ to enhance resolution within the densely populated region.

\section{Experiments}

\subsection{Dataset}
We test our model on a synthetic dataset with intrinsically quantum data, a synthetic dataset of classical origin that is designed to be compressible, and a large and noisy real-world gene expression dataset (including schizophrenia patients and controls) from the PsychENCODE project \cite{wang2018comprehensive}.
For the datasets from a classical data source, we first convert them into density matrices (representing each data point by a pure state) before applying the $\zeta$-QVAE. This allows the $\zeta$-QVAE to be tested on a variety of quantum data sources while enabling direct comparison with classical models.
Further, it should be noted that while we test the performance of our model on a real-world gene expression dataset, this is intended only as a test-bed for the $\zeta$-QVAE, as processing real-world genomics data would require a higher-dimensional architecture.

\subsubsection{Synthetic quantum dataset}
\label{sec:syn_quantum_input}

The synthetic quantum dataset comprises 1000 two-qubit states. To construct the dataset, we first define the preparation state. For the first qubit, we generate random density matrices with a norm of the expectation values of the Pauli $X, Y$ and $Z$ operators between 0.6 and 0.7, i.e., the data points are distributed within a spherical shell inside the Bloch sphere. The second qubit of all data points in the preparation state is initialized to the zero state $\ket{0}$. Next, we apply a Controlled-RY gate with the first qubit as the control. The RY gate acting on the second qubit is parameterized by a random Gaussian variable $\theta\sim\mathcal{N}(\mu=\frac{\pi}{2},\sigma=\frac{\pi}{20})$. The resulting quantum states after applying the Controlled-RY gate form the dataset.
This synthetic quantum dataset is specifically designed to exhibit a data structure with no classical analog while ensuring that the data live on a lower-dimensional manifold, making it inherently compressible.

\begin{figure}[!h]
\centering
\includegraphics[width=0.8\linewidth]{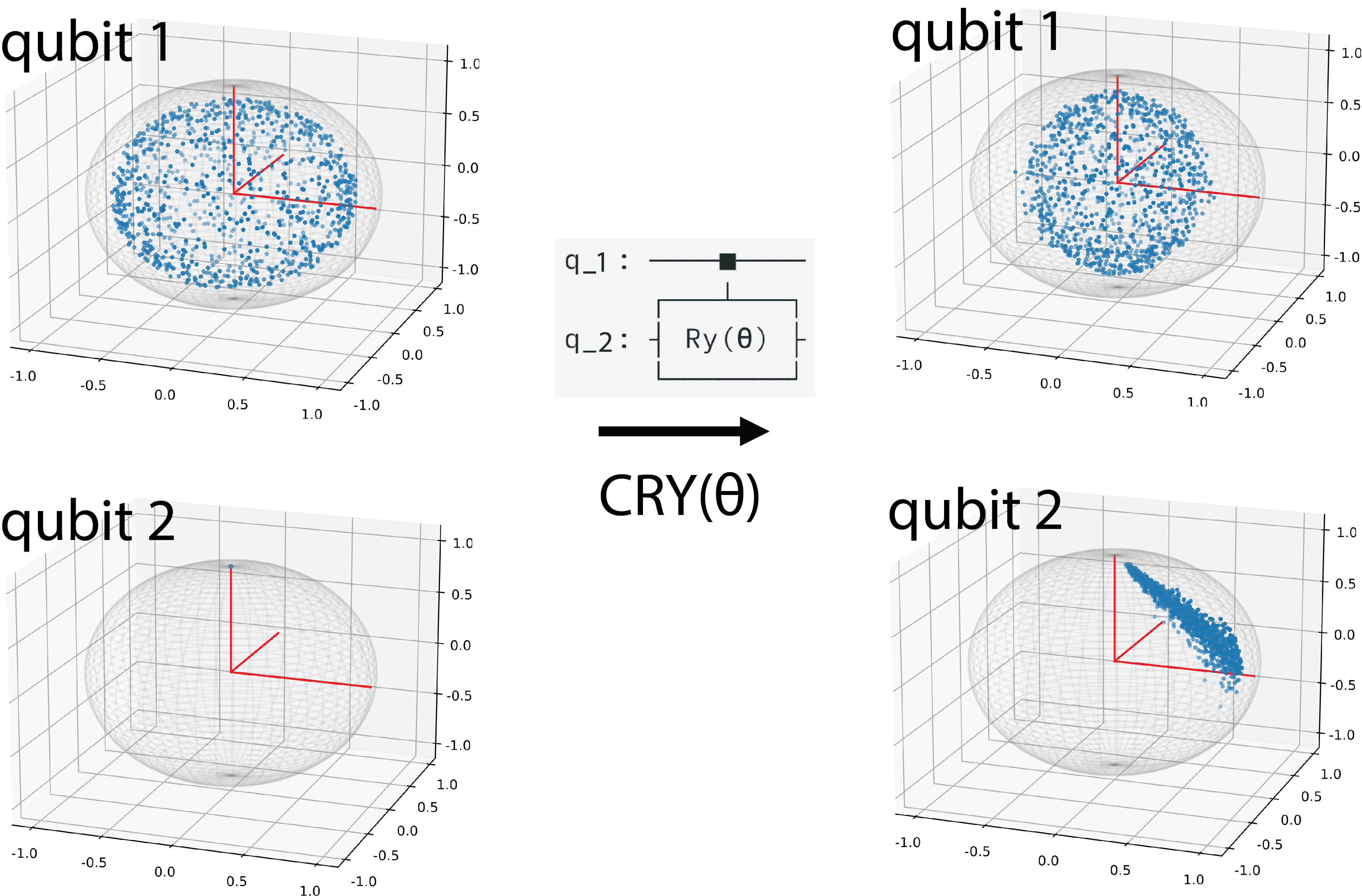}
\caption{The synthetic quantum dataset.}
\label{fig:synInput}
\end{figure}

This process and the distribution of the quantum states are illustrated in \autoref{fig:synInput}. We note that, for visualization, the coordinates $(x,y,z)$ 
of the quantum states are derived from the expectation values of the Pauli operators $(X,Y,Z)$ acting individually on each qubit, i.e., $P_1\otimes I_2$ and $I_1\otimes P_2$, $P \in \{X,Y,Z\}$. Consequently, these visualizations do not capture the correlations between the two qubits.

\subsubsection{PsychENCODE gene expression data}
In this dataset, the schizophrenia status for patients and controls is given together with the quantile normalized expression values of 16 selected genes, generated from RNAseq data from the prefrontal cortex of $\sim1500$ postmortem subjects from the PsychENCODE consortium \cite{wang2018comprehensive}.  These genes were selected from a panel of 555 genes, including pre-identified high-confidence schizophrenia genes and transcription factors.  The 16 genes selected were those found to have the highest variance across patients.

To allow for possible future applications of angle embedding on this dataset, we first conducted a global normalization on the data $$ \vec{x}_i\rightarrow\pi\times[\vec{x}_i-\min(X)]/[\max(X)-\min(X)],$$ where $X$ is the entire dataset matrix (including all feature vectors from all data points) and $\vec{x}_i$ the gene expression vector of the $i$-th data point. We employed amplitude embedding in this study; we therefore performed an additional per-data-point L2 normalization, which allowed us to process the data as state vectors. We randomly picked equal number of cases (patients) and controls to create a training and test partition of size 695 and 298 respectively.

\subsubsection{Swiss Roll synthetic dataset}
In the interest of understanding the generalizability to different datasets, we run the $\zeta$-QVAE and QSVC classifier on $1000$ data points from the Swiss Roll dataset \cite{marsland2014} as implemented in Python's \textit{scikit-learn} package. The Swiss Roll dataset involves a helically distributed sheet of 3-dimensional points that can be compressed to a 2-dimensional manifold. To adapt the dataset to our context, we take the 3-dimensional dataset and append 5 additional dimensions by adding Gaussian-distributed noise terms (zero-centered, standard deviation $=0.2$).  We applied a per-data-point L2 normalization to make the inputs suitable for the quantum circuit. Furthermore, we set up a classification task by designating approximately half of the points as ``cases" and the other half as ``controls"; the task is designed to allow perfect classification along the 2-dimensional manifold, thus serving to evaluate how well we capture the 2-dimensional manifold. 

\begin{figure}[!h]
\centering
\includegraphics[width=0.48\linewidth]{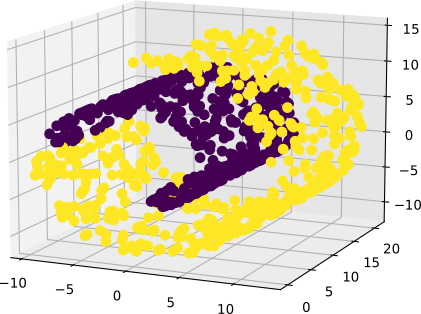}
\caption{The 3-dimensional Swiss Roll dataset. The colors indicate the labels for the classification task.}
\end{figure}

\subsection{Model setup}
To train the $\zeta$-QVAE, we used the COBYLA optimizer with the following configuration: a training duration of 60 epochs and a patience setting of 20 epochs for the PsychENCODE and Swiss roll datasets, and 20 epochs with a patience setting of 5 epochs for the synthetic quantum dataset.
Additionally, we kept the number of layers $N_l$ identical within both the encoder and decoder, as well as the count of auxiliary qubits in the encoder and decoder ($N_A$ and $N_B$). When running the $\zeta$-QVAE with $N_T$ trash qubits, we select the first $N_T$ qubits as trash qubits.
All the performance results, including $\zeta$-QVAE reconstruction rate and QSVC classification accuracies are averaged over five random initializations.

We determined the number of layers of the QSVC classifier $N_{cl}$ based on its performance on the input datasets. Notably, we observed that varying $N_{cl}$ between one and three had negligible impact for the datasets comprising 16 input features. Nevertheless, to account for potential larger input feature dimensions, where a greater $N_{cl}$ might be essential, we opted to set $N_{cl}=3$ for the remainder of the study. Throughout this study, the test accuracy serves as the metric for the classification performance.

\section{Results}
In this section, we thoroughly evaluate the 
$\zeta$-QVAE framework using the instance level objective function in \autoref{sec:objectivefunction_choice}, \ref{sec:sec5_regularization}, \ref{sec:sec5_classification}, \ref{sec:sec5_synQuantum}, and \ref{sec:sec5_swissroll}, while in \autoref{sec:sec5_global} we test the framework using the global level objective function defined on probabilistic mixtures of input data.
We begin by \autoref{sec:objectivefunction_choice} offering an overview of different objective functions introduced in \autoref{sec:training_objectives}. Subsequently, we focus on the specific case of our model in which the negative fidelity serves as the reconstruction loss, complemented by the JSD as the regularization loss. We will refer to this specific objective function as \textit{Fid+JSD} in the following.
In \autoref{sec:sec5_regularization} and \autoref{sec:sec5_classification}, we provide a thorough investigation of the impact of the model architecture on the regularization, and consequently, on the quantum state reconstruction and downstream classification tasks using the \textit{Fid+JSD} objective function.
In \autoref{sec:sec5_synQuantum} and \autoref{sec:sec5_swissroll}, we evaluate the 
$\zeta-$QVAE on two types of inherently compressible synthetic datasets: one derived from an intrinsically quantum origin and the other from a classical origin. Finally in \autoref{sec:sec5_VAE}, we compare the $\zeta-$QVAE with QAE and classical VAE models.

In our model, the architecture is controlled by several hyperparameters, including the number of layers $N_l$ in the encoder and decoder, the $\beta$-value and the number of auxiliary qubits in the encoder and decoder $N_A=N_B$. As we show, the impacts of these hyperparameters are not independent from each other. We evaluate the performance of each model using the fidelity reconstruction rate of the $\zeta$-QVAE and the accuracy of the QSVC on the downstream classification tasks.
The notation employed in this section is as follows: $f(N_A, N_l)$ represents the fidelity reconstruction rate of a given model with $N_A$ auxiliary qubits and $N_l$ layers. Similarly, $l(N_A, N_l)$ signifies the QSVC test accuracy using the latent states of the corresponding model with $N_A$ auxiliary qubits and $N_l$ layers as input, while $r(N_A, N_l)$ denotes the QSVC test accuracy using the reconstructed states as input.

\subsection{Objective function choice:}
\label{sec:objectivefunction_choice}
\textbf{Choice of reconstruction loss}
We begin with the evaluation of the three different forms of reconstruction loss - fidelity, Wasserstein and JSD - at $\beta=0$, which implies that the regularization term is excluded from the objective function. The results are presented in \autoref{tab:variousRecon_full_QVAE}. 

\begin{table}[h]
\centering
\caption{Comparison of the three types of reconstruction losses at $\beta=0$.}
\scalebox{0.8}{
\begin{tabular}{c c c c }
\toprule
&fidelity & wasserstein& JSD \\
$f(0,3)$ & $0.844 \pm0.043$  & $0.853 \pm0.033$ & $0.839 \pm0.037$ \\
$l(0,3)$ &$0.647\pm 0.01$  &  $0.652 \pm0.007$&$0.652\pm 0.013$  \\
$r(0,3)$ & $0.644 \pm0.01$ & $0.65 \pm0.005$ &$0.653 \pm0.012$ \\
\midrule
$f(0,2)$ & $0.898 \pm0.008$ & $0.882 \pm0.013$&$0.875 \pm0.012$  \\
$l(0,2)$ & $0.636 \pm0.008$&$0.65 \pm0.009$ & $0.653 \pm0.011$  \\
$r(0,2)$ &$0.637\pm0.008$ &$0.65 \pm0.01$ &$0.654 \pm0.007$\\
\midrule
$f(0,1)$ & $0.953\pm0$ & $0.771\pm 0.223$ & $ 0.953 \pm0.0$\\
$l(0,1)$ & $0.606 \pm 0.002$ & $0.62\pm 0.019$&$0.605\pm 0.002$  \\
$r(0,1)$ &$0.607\pm0.003$ & $0.622 \pm0.023$ &$0.606 \pm0.005$\\
\midrule
$f(1,3)$ &$0.742 \pm0.046$ &$0.695 \pm0.024$ &$0.686 \pm0.048$ \\
$l(1,3)$ & $0.655\pm 0.003$ &$0.65 \pm0.008$ &$0.651 \pm0.005$\\
$r(1,3)$ &$0.627 \pm0.021$ &$0.607 \pm0.026$ &$0.63 \pm0.016$ \\
\midrule
$f(1,2)$ &   $0.851 \pm0.029$& $0.838\pm 0.026$ & $0.841 \pm0.03$ \\
$l(1,2)$ & $0.653 \pm0.013$ &$0.657 \pm0.011$ & $0.653 \pm0.014$  \\
$r(1,2)$ & $0.613 \pm0.019$ & $0.601 \pm0.023$ & $0.622 \pm0.019$ \\
\midrule
$f(1,1)$ & $0.886 \pm0.009$ & $0.894 \pm0.005$ & $0.9 \pm0.007$  \\
$l(1,1)$ & $0.629 \pm0.018$ & $0.64 \pm0.027$& $0.636 \pm0.023$\\
$r(1,1)$ & $0.574 \pm0.021$ & $0.573 \pm0.03$& $0.571 \pm0.032$\\
\bottomrule
\end{tabular}
}
\label{tab:variousRecon_full_QVAE}
\end{table}

\noindent For $f(0,1)$, the Wasserstein reconstruction loss failed to converge during training in two out of five independent runs. This resulted in a lower average reconstruction rate and a higher standard deviation. For the other settings, we found that all three types of reconstruction loss behave qualitatively similarly at $\beta=0$, which can be briefly summarized as follows (further details are elaborated in \autoref{sec:sec5_regularization} and \autoref{sec:sec5_classification}): 
\begin{itemize}
    \item In the case no auxiliary qubits are employed in the decoder, the reconstructed state is identical to the latent state (extended by reference $\ket{0}$ trash-qubits) up to a unitary transformation and thus results in substantially the same fidelity-based quantum kernel matrix for QSVC. Thus, the classification performance on the latent and reconstructed states is observed to be effectively the same.

    \item Regardless of the presence of auxiliary qubits, with an increasing number of layers, the fidelity reconstruction rate decreases while the test accuracy of classification tasks improves for both latent and reconstructed states. This observation implies 
    that increasing the number of layers may implicitly regularize the model, since the larger parameter search space increases the difficulty for the COBYLA optimizer of finding solutions with high reconstruction fidelity.
    
    \item For any given value of $N_l$, incorporating auxiliary qubits results in a lower fidelity reconstruction rate compared to the case where no auxiliary qubits are used. However, the classification accuracy on the latent states improves slightly while the classification accuracy on the reconstructed states drops especially for smaller $N_l$.
\end{itemize}

\noindent We observe that, generally, the negative fidelity loss is able to achieve better reconstruction performance, while performing comparably to the other losses on classification tasks; we therefore use fidelity reconstruction loss in the following sections.

\textbf{Choice of regularization loss:}
Next, we examine the different choices of regularization loss choices at different values of $\beta$. The results are shown in \autoref{tab:regularization_loss_comparison}.
Given the variations in overall scale among the different forms of regularization loss, our focus shifts to slightly different ranges of $\beta$ for each.
Recalling the baseline results at $\beta=0$ from \autoref{tab:variousRecon_full_QVAE}, the performance of the fidelity reconstruction loss is as follows: $f(0,3)=0.844\pm0.043$, $l(0,3)=0.647 \pm0.01$ and $r(0,3)=0.644 \pm0.01$.

\begin{table}[h]
\centering
\caption{Negative fidelity reconstruction loss with three different regularization loss options.}
\begin{subtable}[c]{1\textwidth}
\centering
\scalebox{0.65}{
\begin{tabular}{c  c  c c c c c c  }
\toprule
 & $\beta=0.5$  &$\beta=1$  & $\beta=1.5$  & $\beta=2$ & $\beta=2.2$ & $\beta=2.5$ & $\beta=2.7$\\
$f(0,3)$&  $0.876\pm0.025$  &$0.843 \pm 0.02$ &$0.813\pm 0.017$&$0.815 \pm0.013$&$0.763 \pm0.043$&$0.765\pm 0.038$ &$0.728 \pm0.035$ \\
$l(0,3)$ &$ 0.651\pm0.007$ &$0.651 \pm 0.011$& $0.654 \pm0.012$&$0.653 \pm0.009$  &$0.658 \pm0.005$ & $0.669 \pm0.005$ &$0.661 \pm0.011$ \\
$r(0,3)$  & $0.653 \pm0.005$  &$0.649 \pm 0.009$&$ 0.652\pm 0.01$  &$0.655 \pm0.01$  &$0.661\pm 0.005 $&$0.665 \pm0.003$ &$0.061\pm0.011$ \\
\bottomrule
\end{tabular}
}
\caption{JSD regularization loss}
 \end{subtable}

 \begin{subtable}[c]{1\textwidth}
  \centering
 \scalebox{0.65}{
\begin{tabular}{c  c  c c c c c }
\toprule
& $\beta=0.5$  &$\beta=1$ &$\beta=1.5$&$\beta=2$&$\beta=2.2$&$\beta=2.5$ \\
$f(0,3)$  &$0.773 \pm0.039$&$0.714 \pm0.037$& $0.664 \pm0.064$&$0.522 \pm0.085$&$0.53 \pm0.076$&$0.425 \pm0.068$ \\
$l(0,3)$ & $0.654 \pm0.014$&$0.653 \pm0.012$& $0.652 \pm0.009$& $0.653 \pm0.014$& $0.656 \pm0.009$&$0.661 \pm0.007$ \\
$r(0,3)$ &$0.654\pm0.011$& $0.652\pm0.012$& $0.652\pm0.008$ &$0.65\pm0.014$&$0.658\pm 0.014$ & $0.661 \pm0.007$ \\
\bottomrule
\end{tabular}
}
\caption{KLD regularization loss}
 \end{subtable}

 \begin{subtable}[c]{1\textwidth}
  \centering
\scalebox{0.65}{
\begin{tabular}{c  c  c c c c c c }
\toprule
& $\beta=0.5$  &$\beta=1$&$\beta=1.1$&$\beta=1.2$ &$\beta=1.5$&$\beta=2$&$\beta=2.5$\\
$f(0,3)$ & $0.863\pm0.009$&$0.864\pm0.025$& $0.847 \pm0.015$&$0.849 \pm0.015$ &$0.834\pm0.027$&$0.829\pm0.033$&$0.798\pm0.03$\\
$l(0,3)$  &$0.648\pm0.007$&$0.646\pm0.015$& $0.657\pm0.008$&$ 0.661\pm0.01$&$0.655\pm0.005$&$0.653\pm0.008$&$0.648\pm0.019$\\
$r(0,3)$ & $0.649\pm0.007$&$0.646\pm0.014$&$0.658 \pm0.01$&$ 0.658\pm 0.01$ &$0.657\pm0.006$&$0.656\pm0.01$& $0.651\pm0.018$  \\
\bottomrule
\end{tabular}
}
\caption{Negative fidelity regularization loss}
 \end{subtable}
 \label{tab:regularization_loss_comparison}
\end{table}

\noindent Here, we observe that KLD has the least favorable performance among the regularization loss options in terms of reconstruction rate.
All three regularization loss options seem to be comparable in classification test accuracy, although the model utilizing JSD performs is slightly better for the optimal $\beta$.  Consequently, our focus in the next section is on the combination of negative fidelity reconstruction loss and JSD regularization loss.

\subsection{Understanding the determinants of regularization and reconstruction in $\zeta-$QVAE}
\label{sec:sec5_regularization}
Models with appropriately tuned regularization, leading to an optimal degree of disentanglement in their latent representations, have been shown to outperform those lacking such adjustments due to their ability to capture the independent underlying latent factors effectively \cite{higgins2016beta}.
In this section, we investigate how the degree of regularization (explicit and implicit, as discussed below) and the reconstruction rate of the $\zeta-$QVAE are influenced by the interplay of several factors: the $\beta$-value, the presence of auxiliary qubits and the circuit complexity. 
We impose different circuit complexity constraints by varying the number of layers $N_l$ in the encoder and decoder and studied a range of $\beta$-values from zero to six, while considering all combinations with and without one auxiliary qubit. 

The reconstruction ability of the model is estimated using the fidelity reconstruction rate.
To quantify the regularization effect, we analyze the distribution of the latent states in the latent space by calculating the regularization loss.  In addition, we take into account downstream classification performance as an additional metric for the evaluation of the reconstruction rate and degree of regularization. 

\begin{table}[h]
\centering
\caption{\textit{Fid+JSD} objective function}
\scalebox{0.8}{
\begin{tabular}{c c c c }
\toprule
&$\beta=0$ & $\beta=1$ &$\beta=2$\\
$f(0,3)$ & $0.844 \pm0.043$&$0.843 \pm0.02$& $0.815\pm 0.013$\\
$l(0,3)$ &$0.647 \pm0.01$  &  $0.651 \pm0.011$&$0.653 \pm0.009$\\
$r(0,3)$ & $0.644 \pm0.01$ & $0.649 \pm0.009$ &$0.655 \pm0.01$\\
\midrule
$f(0,2)$ & $0.898 \pm0.008$ & $0.883 \pm0.018$ &$0.832 \pm0.027$\\
$l(0,2)$ & $0.636 \pm0.008$ &$0.644\pm 0.009$ &$0.653 \pm0.009$ \\
$r(0,2)$ & $ 0.637  \pm0.008$ &$0.644 \pm 0.004$ &$0.655 \pm 0.007$\\
\midrule
$f(0,1)$ & $0.953 \pm0.0$ & $0.673 \pm0.212$ &$ 0.728 \pm0.186$\\
$l(0,1)$ & $0.606 \pm0.002$ & $0.646 \pm0.006$ &$0.653 \pm0.008$ \\
$r(0,1)$ & $0.607 \pm0.003$ &$0.647\pm 0.003$ &$0.653 \pm0.01$ \\
\midrule
$f(1,3)$ & $0.742\pm0.046$ &$0.649 \pm0.061$  &$0.561 \pm0.066$\\
$l(1,3)$ & $0.655\pm 0.003$ &$0.651 \pm0.011$ &$0.651 \pm0.016$\\
$r(1,3)$ &$0.627 \pm0.021$ & $0.613\pm 0.004$ &$0.605 \pm0.011$\\
\midrule
$f(1,2)$ &   $0.851 \pm0.029$& $0.815 \pm0.03$ &$0.686 \pm0.044$\\
$l(1,2)$ & $0.653 \pm0.013$ & $0.638 \pm0.01$ &$0.644 \pm0.005$\\
$r(1,2)$ & $0.613 \pm0.019$ & $0.6 \pm0.027$ &$ 0.593 \pm0.024$\\
\midrule
$f(1,1)$ & $ 0.886\pm 0.009$ & $0.887 \pm0.009$ &$0.887 \pm0.006$ \\
$l(1,1)$ & $0.629 \pm0.018$&  $0.648 \pm0.005$ &$0.648 \pm0.012$\\
$r(1,1)$ & $0.574 \pm0.021$ & $0.588 \pm0.014$ &$0.562 \pm0.022$\\
\bottomrule
\end{tabular}
}
\label{tab:FidJSD}
\end{table}

We noticed that the models with $N_l=1$ and no auxiliary qubits often failed to converge at non-zero $\beta$ values, leading to the large standard deviation of $f(0,1)$ in \autoref{tab:FidJSD}.  For example, among the five random initializations at $\beta=2$, two exhibited a test fidelity reconstruction rate around $0.5$ while the remaining three had a fidelity of approximately 0.88. Similarly, at $\beta=1$, three had a fidelity around 0.5, and the remaining two showed a fidelity near 0.93. This may be due to the limited number of free parameters in the model when only a single layer is used.
Further, for the models with $N_l=1$ and one auxiliary qubit, we found that for the range of $\beta \in [0,6]$, the reconstruction fidelity remained constant within the error range as suggested by $f(1,1)$ in \autoref{tab:FidJSD}. 
Below we provide several key findings based on results obtained for $N_l=2$ and 3.

\begin{figure}[h!]
\centering
\includegraphics[width=0.7\linewidth]{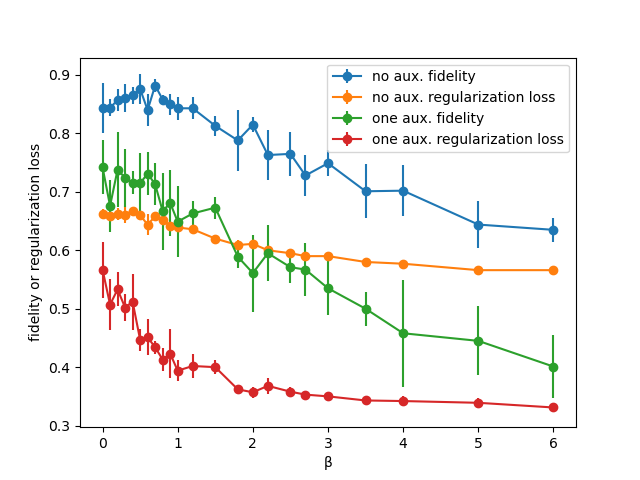}
\caption{The two components of the objective function are plotted as a function of $\beta$ for the case with one trash qubit and $N_l=3$.}
\label{fig:fid_regu_beta}
\end{figure}

\textbf{Regularization is controlled by $\beta$-value, model complexity and number of auxiliary qubits}: 
Varying $\beta$ is the most direct way to adjust the degree of the regularization.
In \autoref{fig:fid_regu_beta}, the fidelity reconstruction rate (which is 1 for a perfect reconstruction) and regularization loss (where 0 implies stronger regularization / smaller regularization loss) are shown as a function of $\beta$. We see that for the entire range of $\beta$ considered in this study, higher $\beta$ values lead to stronger regularization and worse fidelity reconstruction rates. 
One can also see in \autoref{fig:fid_regu_beta}, that the regularization loss is smaller when using one auxiliary qubit compared to the scenario without any auxiliary qubits, across all values of $\beta$.
\noindent In addition, the model complexity controlled by $N_l$ can also influence the degree of the regularization.
As we can see in \autoref{tab:FidJSD} in the case where no auxiliary qubits are present, similar to our observation at $\beta=0$, increasing the number of layers in the encoder and decoder also leads to a decreased reconstruction rate for $\beta=1$ and 2, accompanied by improved classification performance. Increasing the number of layers or the number of auxiliary qubits thus has a similar effect to increasing $\beta$, resulting in a form of implicit regularization as noted above in \autoref{sec:objectivefunction_choice}. At $\beta=0$, the effect of increasing number of layers is more noticeable than at a higher value of $\beta$.
The number of layers, auxiliary qubits and $\beta$ can thus be viewed as jointly contributing to the regularization of the model.

\begin{figure}[h!]
\begin{subfigure}{0.48\textwidth}
\includegraphics[width=1\linewidth]{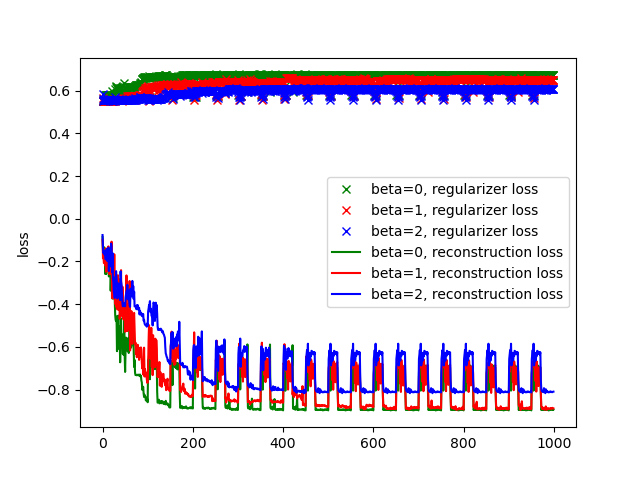}
\caption{$N_l=2$, no auxiliary qubits}
\end{subfigure}
\begin{subfigure}{0.48\textwidth}
\includegraphics[width=1\linewidth]{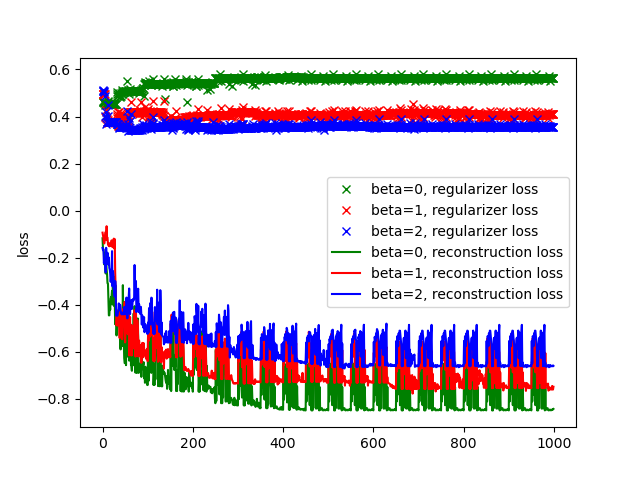}
\caption{$N_l=2$, one auxiliary qubit}
\end{subfigure}
\begin{subfigure}{0.48\textwidth}
\includegraphics[width=1\linewidth]{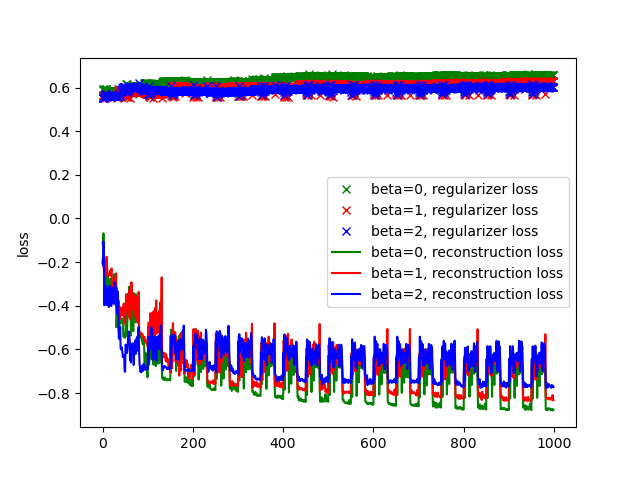}
\caption{$N_l=3$, no auxiliary qubits}
\end{subfigure}
\begin{subfigure}{0.48\textwidth}
\includegraphics[width=1\linewidth]{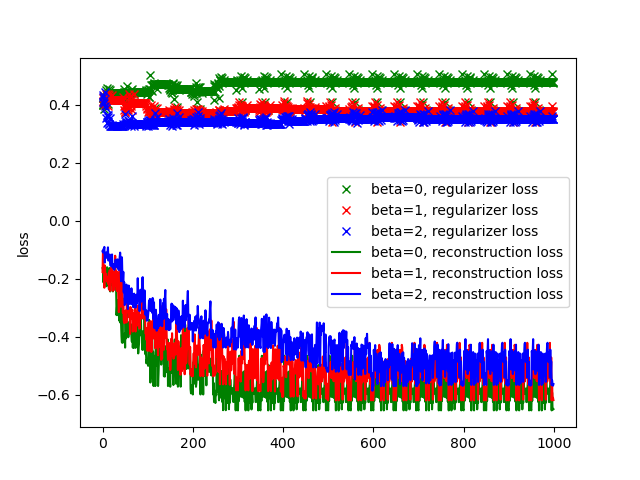}
\caption{$N_l=3$, one auxiliary qubit}
\end{subfigure}
\caption{The optimization process of the reconstruction loss and regularization loss are shown separately. In the absence of auxiliary qubits, the two components of the objective function are mutually dependent. Utilizing auxiliary qubits allows for decoupling and thus simultaneous improvement of both terms. The fluctuations in the curves (e.g., repeated upward jumps of the reconstruction loss) are caused by the large learning rate during the initial iterations of each new epoch.}
\label{fig:beta_noAux_vs_Aux_separate_optimization}
\end{figure}

\textbf{Reconstruction and regularization are strongly dependent in the absence of auxiliary qubits:} 
In the case where no auxiliary qubits are used, the reconstructed states are obtained from the latent states by a unitary (linear) operation. 
This means effectively, both the reconstruction constraints and the regularization constraints are imposed to the same space (since the latent states are mapped to a linear subspace of the output space with the same intrinsic dimensionality). The reconstructed state thus inherits directly the same regularization as the latent state. This can be seen in the left panels in \autoref{fig:beta_noAux_vs_Aux_separate_optimization}, where a lower reconstruction loss at smaller $\beta$ can only be achieved by sacrificing the regularization loss, $\ie\,$ by allowing a higher regularization loss.

\textbf{Reconstruction and regularization are substantially decoupled in the presence of auxiliary qubits:} 
In the presence of auxiliary qubits, we observed simultaneous decreasing curves for the regularization loss and the reconstruction loss in right panels of \autoref{fig:beta_noAux_vs_Aux_separate_optimization} (one should note that the reconstruction loss is the negative counterpart of the reconstruction fidelity shown in \autoref{fig:fid_regu_beta}).
This is because the presence of auxiliary qubits allows for non-unitary transformations from latent states to reconstructed states, thus allowing a separate optimization of the regularization loss and reconstruction loss. This feature of $\zeta-$QVAE has no classical analogue and can be potentially utilized to devise a framework for controlling the degree of coupling between latent and reconstructed states, thus enabling a flexible trade-off between regularization and reconstruction rates.

\textbf{Reconstructing the original state poses challenges in the presence of auxiliary qubits:} 
As shown in \autoref{tab:FidJSD}, the fidelity reconstruction rate is lower in the presence of one auxiliary qubit compared to its absence. In addition, we note a decrease in the classification performance on the reconstructed states when one auxiliary qubit is used compared to when no auxiliary qubits are used. Whereas for the latent states, the classification performance remains similar. 
The observed phenomenon may be explained by the removal of the constraint imposed by the coupled reconstruction loss and regularization loss. While adding an auxiliary qubit alleviates this constraint, it also introduces a greater challenge to the optimization process of the model parameters\cite{ragone2023representation}. This difficulty in optimization may be exacerbated by the Barren plateau effect intensified by the inclusion of an additional qubit \cite{Arrasmith2021}.

\subsection{Optimal representations for downstream classification tasks}
\label{sec:sec5_classification}
\begin{figure}
\begin{subfigure}{0.48\textwidth}
\includegraphics[width=1\linewidth]{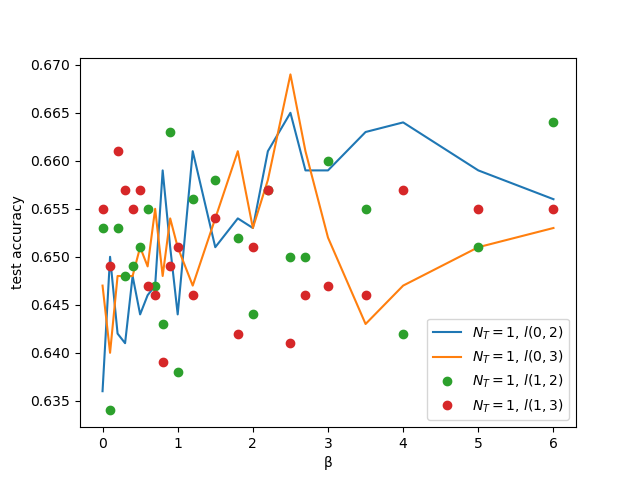}
\caption{$N_T=1$}
\end{subfigure}
\begin{subfigure}{0.48\textwidth}
\includegraphics[width=1\linewidth]{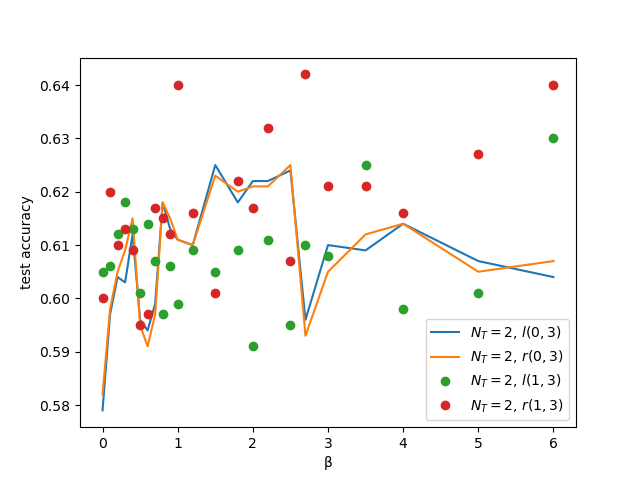}
\caption{$N_T=2$}
\end{subfigure}

\begin{subfigure}{0.48\textwidth}
\includegraphics[width=1\linewidth]{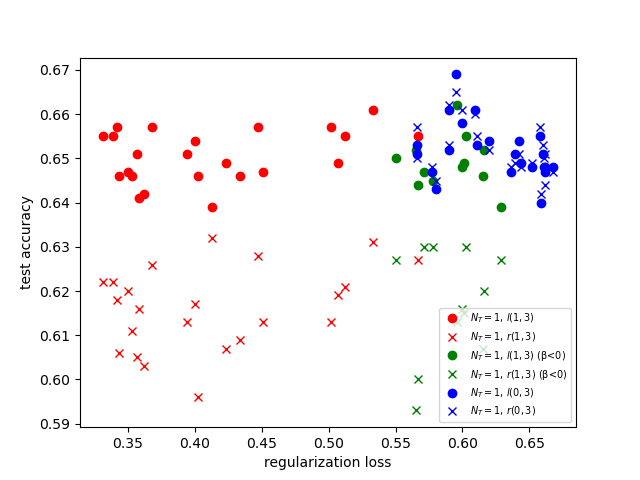}
\caption{$N_T=1$, including negative $\beta$ in green}
\end{subfigure}
\begin{subfigure}{0.48\textwidth}
\includegraphics[width=1\linewidth]{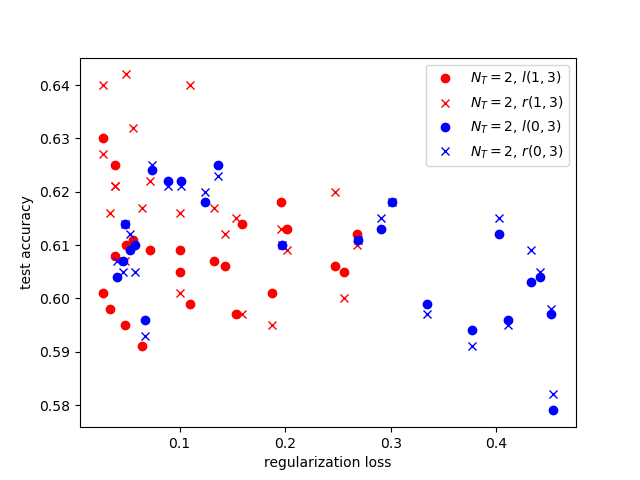}
\caption{$N_T=2$}
\end{subfigure}

\caption{(a) and (b): Classification performance is plotted against $\beta$. For both $N_T=1$ and 2, the no auxiliary qubit cases (orange and blue line) clearly show an optimal $\beta$ with improved classification performance, while in the one auxiliary qubit case the optimal $\beta$ range is unclear. 
For $N_T=2$ using auxiliary qubits is advantageous compared to no auxiliary qubits. 
(c) and (d): Plots test accuracy directly against regularization loss to eliminate uncertainties caused by the intermediate parameter $\beta$. For both $N_T=1$ and 2, while in the no auxiliary qubit case, there is clearly an optimal range for regularization loss, for the one auxiliary qubit case this is less clear.}
\label{fig:noAux_vs_Aux_fidvsaccu}
\end{figure}

\textbf{An optimal degree of regularization exists for the downstream classification performance:} In \autoref{fig:noAux_vs_Aux_fidvsaccu} (a) and (b), we plot the downstream classification performance against $\beta$. 
We noticed that when no auxiliary qubits are used, 
an optimal range of $\beta$ is associated with higher classification accuracy. 
For the one trash qubit case, $\ie$ $N_T=1$, shown in (a), the optimal $\beta$ is found to be around $2.5$ for both $N_l=2$ and $N_l=3$. We also note that the model using three layers achieved slightly higher classification accuracy than the two-layer model.
On the other hand, when one auxiliary qubit is used, the regularization seems to have no clear impact on the downstream classification performance (red and green points). Nevertheless, we cannot conclude less significant improvements cannot be identified since there is a large range of uncertainty in the performance.
In panel (b) where two trash qubits are used, the optimal $\beta$ occurs around 2 for the scenario without auxiliary qubits. Although it is still difficult to determine if there exists an optimal range of $\beta$ in the scenarios with one auxiliary qubit, we can see that the performance of the models with auxiliary qubits is slightly better than that without auxiliary qubits. 

\noindent In panels (c) and (d), we plot the test accuracy directly against the regularization loss. For both $N_T=1$ and $N_T=2$ without auxiliary qubits, we see a clear optimal range of regularization loss at 0.6 (for $N_T=1$) and 0.11 (for $N_T=2$). 
For models with auxiliary qubits, we included also negative $\beta$s, indicated by green points. This is motivated by the observation that, even for $\beta=0$, the regularization was already stronger than the optimal range observed for cases without auxiliary qubits case (blue points).
As shown in (c), $l(1,3)$ (red points) improves slightly with increasing regularization loss in the $N_T=1$ case, and this trend persists with a slight improvement for negative $\beta$ values. In (d), The data suggests an upward trend as the regularization loss decreases. However, this observed trend is less pronounced compared to models without auxiliary qubits and remains suggestive rather than conclusive.

\textbf{Regularization is more advantageous for smaller latent space:} 
For $N_T=1$, the latent space is an eight-dimensional Hilbert space formed by three qubits while for $N_T=2$, the latent space is four-dimensional formed by two qubits.
As shown in \autoref{fig:noAux_vs_Aux_fidvsaccu} (a) and (b), for the $N_T=1$ case, the regularization improves the classification performance by $\approx 4\%$ while for $N_T=2$, the improvement is over $7.5\%$.

\textbf{Classification performance on the latent states is similar to that on the input states:} 
The classification performance of the employed QSVC on the input states is $0.675\pm0.003$.\footnote{We also tested a classical SVC with RBF kernel on the input states and obtained a classification performance of $0.648\pm0.016$, lower than that of the QSVC. The error range in this case is obtained by averaging over various data partitions.}
For one trash qubit case, $\ie$ compressing to half of the original dimensionality, the best classification performance achieved on the latent states is $0.669 \pm0.005$ for $\beta=2.5$, $N_l=3$ and no auxiliary qubits. Notably, this is only $0.9\%$ lower than that achieved with the full original states.
For two trash qubits, $\ie$ compressing to a quarter of the original dimensionality, we achieved a classification performance of $0.63 \pm0.015$ for $\beta=6$, $N_l=3$ and one auxiliary qubit.

\subsection{Training using global objective}
\label{sec:sec5_global}
Recall from \autoref{eq:def_global_state}, that the global state is defined as a mixed state over the entire input dataset.
In this section, we test the performance of the $\zeta$-QVAE using the global density matrix. We consider only the setup where negative fidelity serves as the reconstruction loss and JSD acts as regularization loss. 

In this scenario, our quantum circuit is trained on a single global input state, while the model construction is identical to that of the instance-level model. Hence, through the training phase, one single latent state and one output state are present.
Following the completion of quantum circuit training, each individual instance-level input data point will be fed through the optimized model. For each data point within the original dataset, the associated latent state and reconstructed state are computed. 
Subsequently, calculations for the fidelity reconstruction rate calculation and downstream classification tasks are executed on the instance-level input, latent and reconstructed states.

\begin{figure}[!h]
\centering
\begin{subfigure}{0.49\textwidth}
\includegraphics[width=1\linewidth]{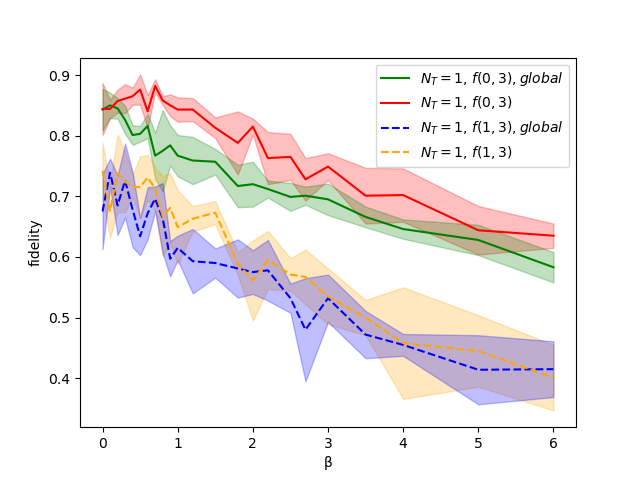}
\caption{Fidelity reconstruction rate}
\end{subfigure}
\begin{subfigure}{0.49\textwidth}
\includegraphics[width=1\linewidth]{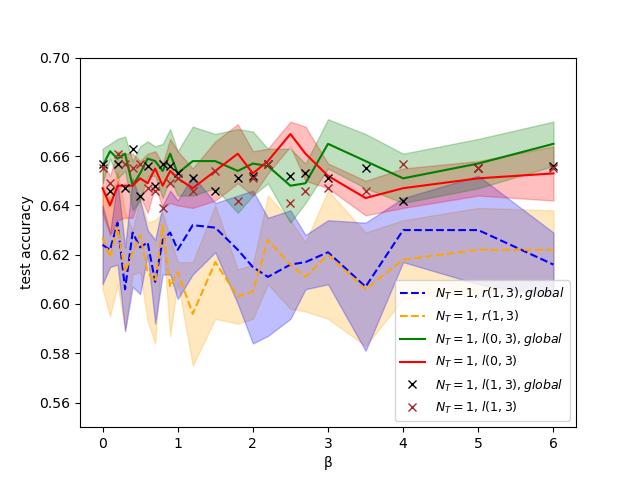}
\caption{Classification performance}
\end{subfigure}
\caption{For $N_T=1$ and $N_l=3$, we compare the globally trained with the instance-level trained model. The shaded areas represent the error range.}
\label{fig:global_test}
\end{figure}

We tested a range of $\beta$s on the global $\zeta-$QVAE and the results are shown in \autoref{fig:global_test}. While the reconstruction rate is slightly lower for nearly all $\beta$s, the overall pattern of the curve with respect to $\beta$ is very similar to that of the instance-level trained models. In the down-stream classification tasks, the QSVC test accuracy achieved on the latent and reconstructed states remains comparable for $\zeta-$QVAE models trained on both global and instance-level data. For $l(0,3)$, where an optimal $\beta$ of 2.5 was observed for the instance-level trained models, the globally trained models exhibit an optimal $\beta$ of three. Nevertheless, the disparity in performance falls within the error range.

\subsection{Application to the synthetic quantum dataset}
\label{sec:sec5_synQuantum}
\begin{figure}[!h]
\centering
\begin{subfigure}{0.57\textwidth}
\includegraphics[width=1\linewidth]{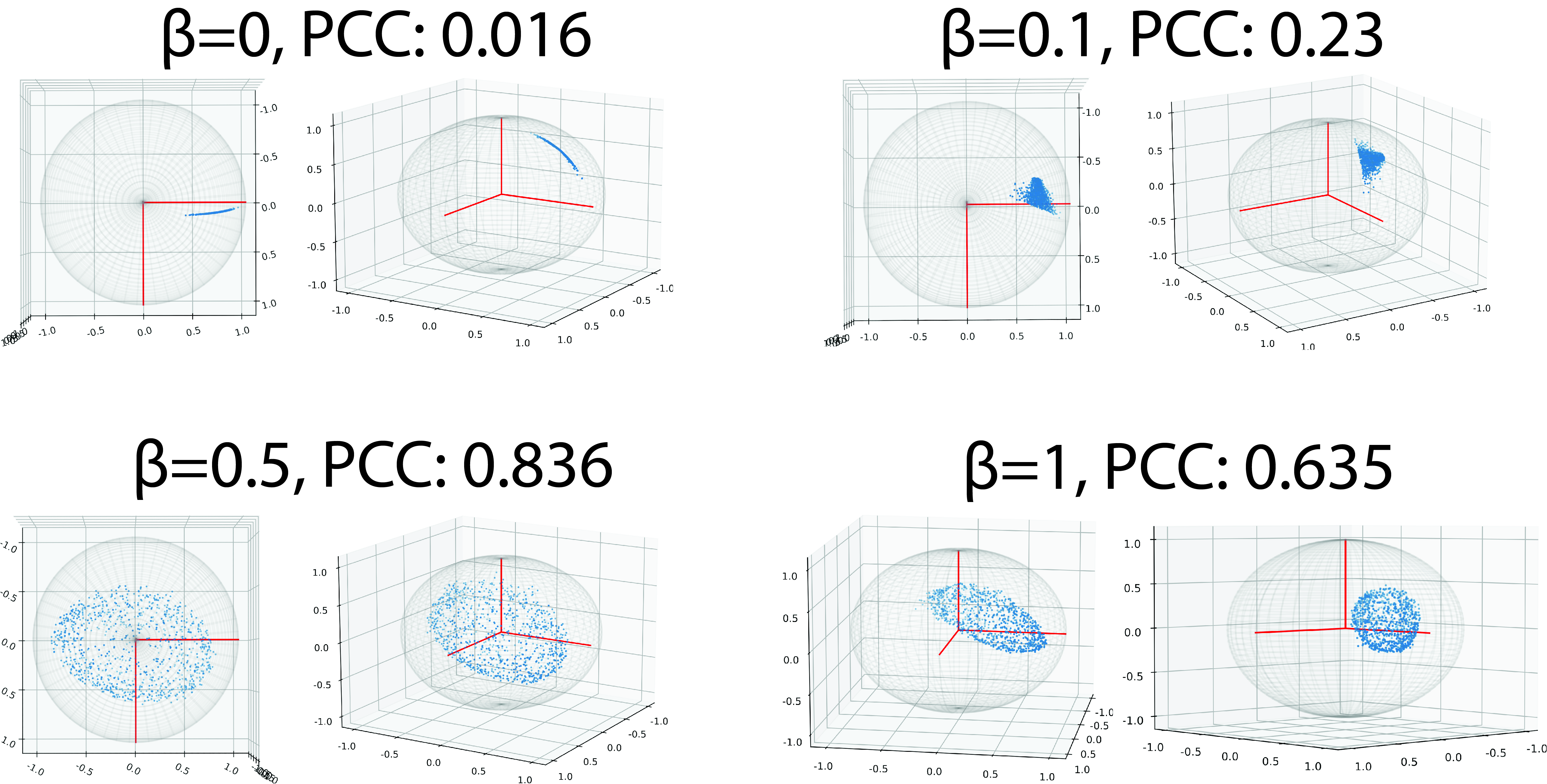}
\subcaption{Two perspectives of the latent Bloch sphere, and input-latent PCC}
\end{subfigure}
\begin{subfigure}{0.42\textwidth}
\includegraphics[width=1\linewidth]{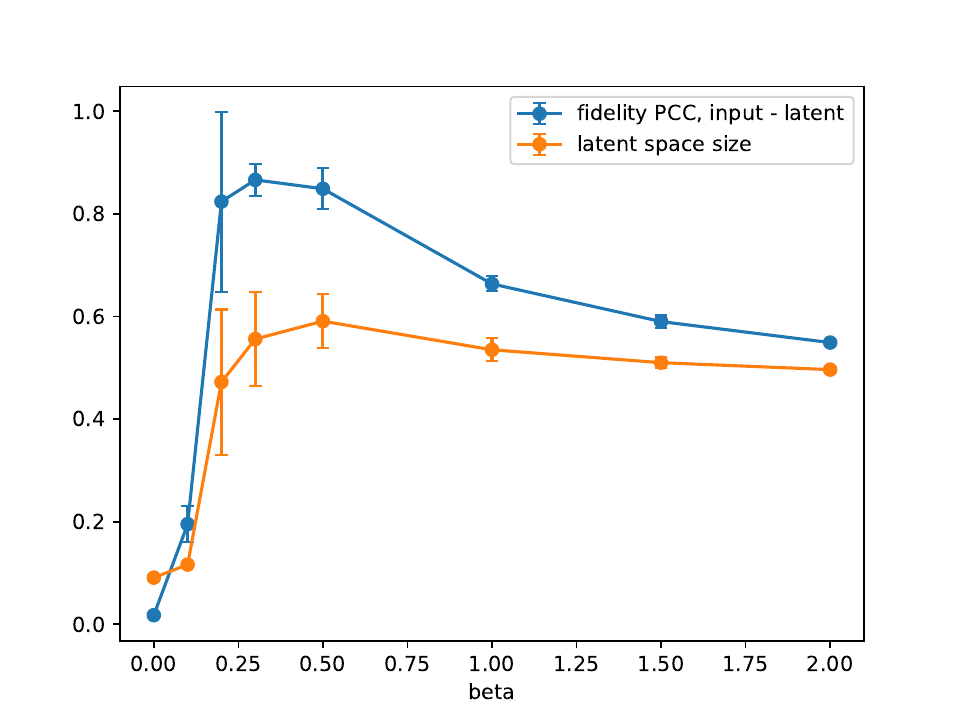}
\subcaption{Input-latent PCC and $\text{Vol}_{\text{latent}}$}
\end{subfigure}
\caption{Latent state features of the synthetic quantum dataset.}
\label{fig:syn_quantum_latent_distribution}
\end{figure}

For this dataset, we compressed the original two-qubit states into a one-qubit latent space, i.e., using one trash qubit ($N_T = 1$), and we learn a $\zeta$-QVAE with a fidelity reconstruction loss and JSD regularization loss. The $\zeta$-QVAE was configured with $N_l = 3$, and we examined  different $\beta$ values: $\beta = {0, 0.1, 0.2, 0.3, 0.5, 1, 1.5, 2}$.
The latent states for various $\beta$ values are visualized within the Bloch sphere from two perspectives in \autoref{fig:syn_quantum_latent_distribution}(a). Their coordinates $(x, y, z)$ are derived from the expectation values of the Pauli operators $(X, Y, Z)$, as described in \autoref{sec:syn_quantum_input}.

In \autoref{fig:syn_quantum_latent_distribution}(b), we provide a quantitative estimate of the model's utilization of the latent space's representation capacity, which is measured as the norm of the standard deviation of the Bloch sphere coordinates of all data points: $\text{Vol}_{\text{latent}}=\sqrt{(\sigma_x^2+\sigma_y^2+\sigma_z^2)}$. 
To evaluate how well the pairwise similarity between the input data points is preserved in the latent space, we compute the quantum fidelity between each pair of data points. We then compute the Pearson correlation coefficient (PCC) between the pairwise fidelity matrices in the input and latent states (input-latent PCC).

We observe that the utilization of the latent space's representation capacity improves with increasing $\beta$ up to an optimal value, beyond which it decreases as $\beta$ is increased further. This is a result of the interplay between reconstruction and regularization. While regularization promotes an isotropic distribution of latent states, too large a value of $\beta$ forces all latent states to gather at the maximally mixed state, i.e., the center of the Bloch sphere.

We also observe a strong positive correlation between $\text{Vol}_{\text{latent}}$ and the input-latent PCC, with a Pearson correlation coefficient of $0.934$ and a p-value of $0.0007$. This suggests that a more effective lower-dimensional latent representation is achieved when a greater portion of the latent space is utilized, which can be optimized by tuning $\beta$ to its optimal value. 

The above suggests that, when using the $\zeta$-QVAE as a generative model, it is important to determine  empirically the distribution of the ensemble of states in the latent space if the goal is to generate data matching a given source at the instance level; while the global density matrix of the output will remain unchanged for any ensemble realizing the maximally mixed state in the latent space, the instance-level characteristics of the output data may differ markedly for different ensembles.

\subsection{Application to the Swiss Roll dataset}
\label{sec:sec5_swissroll}
\begin{figure}[!h]
\centering
\begin{subfigure}{0.48\textwidth}
\includegraphics[width=1\linewidth]{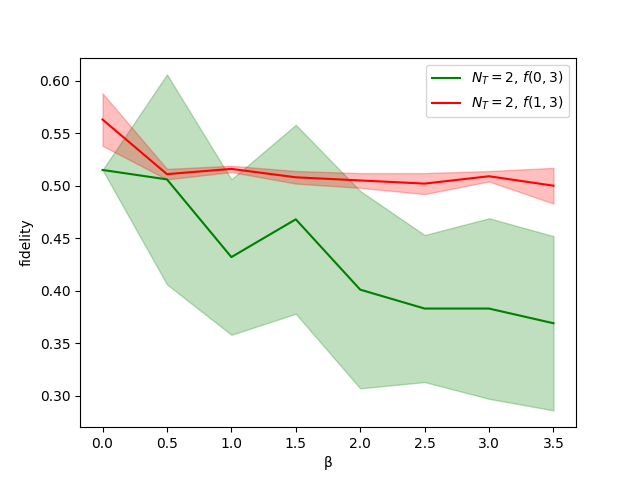}
\caption{Fidelity reconstruction rate}
\end{subfigure}
\begin{subfigure}{0.48\textwidth}
\includegraphics[width=1\linewidth]{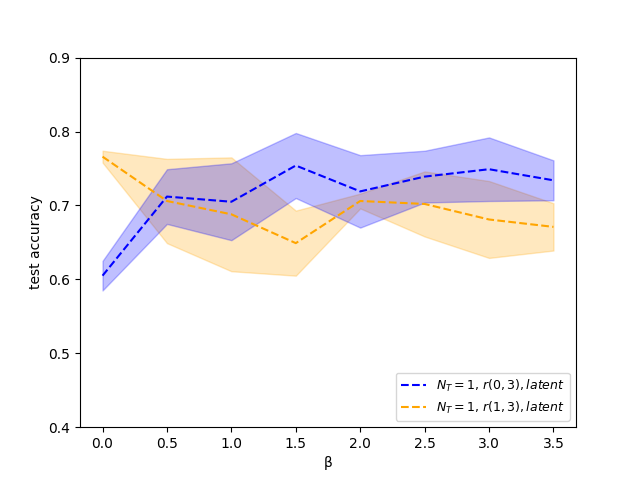}
\caption{Classification performance}
\end{subfigure}
\bigskip

\begin{subfigure}{0.75\textwidth}
\includegraphics[width=1\linewidth]{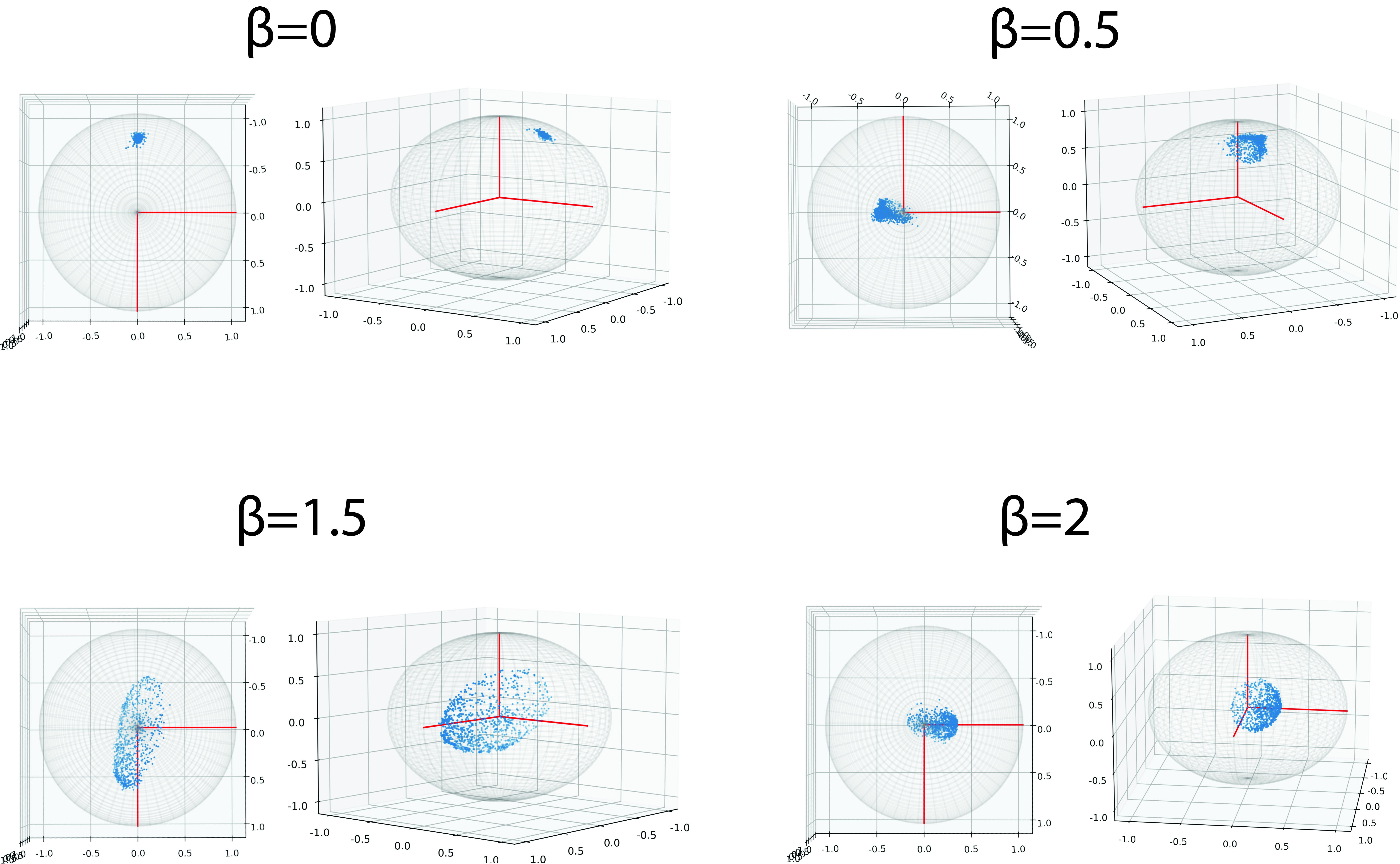}
\caption{Latent states distribution}
\end{subfigure}
\caption{$N_T=2$, $N_l=3$. We evaluate the performance of the $\zeta$-QVAE (a) and QSVC classifier on the latent states (b) using the Swiss Roll dataset. The shaded areas represent the error range. (c) shows two perspectives of the latent representations in Bloch sphere.}
\label{fig:swissroll}
\end{figure}

 To evaluate performance on the Swiss Roll dataset (\autoref{fig:swissroll}), we considered the case where our 8-dimensional input state is mapped to 3 qubits and the latent state is determined by 1 qubit. For the $\zeta$-QVAE, we set $N_T=2$, $N_l=3$, $\beta=\{0,0.5,1.0,1.5,2.0,2.5,3.0,3.5\}$, and test the scenario with zero auxiliary qubits to one with 1 auxiliary qubit added to the encoder and decoder. The results yield similar conclusions to those of the gene expression dataset. The reconstruction fidelity on the leave-out test set steadily decreases with an increase in $\beta$, irrespective of the number of auxiliary qubits (\autoref{fig:swissroll}a). In contrast, the utilization of the latent space's representation capacity increases up to $\beta=1.5$ as visualized in \autoref{fig:swissroll}c.
 The test accuracy on the classification task using the learned latent states achieves a peak at $\beta=1.5$ for the 0-auxiliary-qubits case, showing an improved test accuracy of $0.75 \pm 0.04$ relative to the accuracy of $0.60 \pm 0.02$ at $\beta=0$. We note that the optimal $\beta$ value for the classification task corresponds to the one that maximizes the utilization of the latent space.
 
 On the other hand, there is no clear benefit of a non-zero $\beta$ for the 1-auxiliary-qubit case, at least at the values screened here (\autoref{fig:swissroll}b). This may be due to the fact that the implicit regularization due to the inclusion of the auxiliary qubit is already quite strong at $\beta=0$. Overall, the test accuracy remains reasonably high, with a maximum of $0.75 \pm 0.04$ (based on both the latent and reconstructed states; test AUC = $0.82 \pm 0.05$) for 0 auxiliary qubits and $0.77 \pm 0.01$ (based on the latent state; test AUC = $0.82 \pm 0.02$) for 1 auxiliary qubit.     

\subsection{Comparison to QAE and classical VAEs}
\label{sec:sec5_VAE}
\textbf{Quantum Autoencoder (QAE):}
At $\beta=0$, $\zeta$-QVAE without the regularization term and with a fidelity-based reconstruction loss is similar to the QAE introduced in Ref.\,\cite{romero2017quantum} with the following minor differences: (1) The objective function to be maximized in Ref.\,\cite{romero2017quantum}, $\ie$ fidelity on the trash state $F(\rho_t,\ket{0})$, serves as an upper bound of the actual reconstruction fidelity $F(\rho_i,\mathcal{D}(\mathcal{E}(\rho_i)))$, which we optimize directly; (2) The decoder in QAE is the inverse of encoder, which is a special instance of our decoder, whose parameters are independent from that of the encoder.
Our results show that the $\zeta$-QVAE achieves improved classification performance at $\beta>0$, suggesting that models with regularization offer advantages compared to the QAE.

\noindent\textbf{Classical VAEs:} We compare two types of classical $\beta$-VAEs \cite{higgins2016beta} to the $\zeta-$QVAE.  The first type has a single linear layer without an activation function for both the encoder and decoder. The second type is a two-layer $\beta$-VAE with 12 hidden nodes and a RELU activation function. We consider classical $\beta$-VAEs with 8-, 4-, 2- and 1-dimensional latent spaces. 
Across these cases, we conducted tests over a wide range of $\beta$ and presented the highest classification performance overall $\beta$s in \autoref{tab:classical_VAE}.
It is important to note that in the classical $\beta$-VAE, each dimension in the latent space includes a mean and a variance, resulting in two degrees of freedom per dimension. Therefore, a four/two/one latent dimensional classical VAE is comparable to three/two/one latent qubits in the $\zeta-$QVAE, respectively.

On the gene expression data, the fully quantum compression and classification scheme (QVAE+QSVC) reached a classification accuracy of $0.669\pm0.005$ using three latent qubits and $0.63\pm0.015$ with two latent qubits, outperforming the fully classical compression and classification scheme (VAE+SVC).  On the synthetic Swiss Roll dataset, the $\zeta-$QVAE with one latent qubit achieved a classification accuracy of $0.77\pm0.01$, which is slightly higher than that of the classical $\beta-$VAE with one-dimensional latent space. 

In addition to the improved classification accuracy on the latent states, the number of parameters used by the $\zeta-$QVAE is also much smaller than that of the classical VAE. For example, in the case of 16 input features, a single-layer classical $\beta-$VAE with a 4-dimensional latent space has 216 free parameters. In contrast, a 3-layer $\zeta-$QVAE has only 60 free parameters for the same input features, but benefits from the high dimensionality of the Hilbert space associated with the latent state.  It is also possible that the entanglement between the qubits employed by the model contributes to a reduction in the number of parameters needed for encoding the original data.
Although the quantum scheme offers advantages over the classical approach when the number of model parameters of the classical approach is limited, it is clear that quantum models cannot yet be scaled to address real problem sizes in fields such as genomics. However, the comparison under the constraint of similar model sizes highlights the promising potential of quantum models.

\begin{table}[!h]
\centering
\caption{Classification performance on the latent representations}
\scalebox{1}{
\begin{tabular}{c | c | c | c | c | c }
\toprule
& \multicolumn{2}{c}{gene expression data} & \multicolumn{2}{|c|}{Swiss Roll dataset} \\
& linear VAE& standard VAE & linear VAE& standard VAE \\
\midrule
latent dim.\,= 8 & $0.653\pm0.027$ & $0.659 \pm 0.01$ & $-$ & $-$ \\
latent dim.\,= 4 & $0.643 \pm0.017$ & $0.646 \pm0.006$ & $-$ & $-$ \\
latent dim.\,= 2 & $0.615 \pm0.01$ & $0.612 \pm0.019$& $0.784\pm 0.003$& $0.775\pm 0.006$\\
latent dim.\,= 1 & $-$ & $-$ & $0.708 \pm0.004$& $0.768 \pm0.004$\\
\bottomrule
\end{tabular}
}
\label{tab:classical_VAE}
\end{table}

\section{Advantages of the $\zeta$-QVAE framework}

In general, the application domains of $\zeta-$QVAE are not expected to differ significantly from those of classical VAEs. However, certain distinctive features of $\zeta-$QVAE can offer specific advantages in select applications.

\textbf{Application to large-scale datasets.}
Our framework addresses key challenges in applying quantum models to fields involving large-scale datasets by finding lower-dimensional representations of big datasets with large feature spaces while  maintaining the relationships between data points and preserving essential information crucial for downstream analyses, such as classification. 
This reduces the necessary (quantum) data storage capacity and addresses the limited availability of quantum hardware by allowing subsequent analysis to be carried out by quantum devices with a small number of qubits.
If the original dimensionality is feasible for quantum hardware, the method becomes valuable if the complexity of subsequent analysis is substantially reduced by applying the the $\zeta-$QVAE. In cases where the problem size is too large for quantum hardware, a hierarchical approach based on the $\zeta-$QVAE can be employed. Specifically, the original feature-space can be split by partitioning it into subsets of input features, ideally accounting for correlations within the data. The $\zeta-$QVAE can then be applied to compress each subset of features, producing intermediate compressed data representations. A second round of $\zeta-$QVAE can then be applied to these intermediate states to further reduce dimensionality of the data.

\textbf{Application to privacy-aware computation.}
Our formulation of global objectives holds potential for privacy-preserving computation, as it potentially eliminates the need for access to all the original data points during model training. Instead, only the global density matrix may be required, or alternatively samples may be provided from any equivalent quantum ensemble with the same density matrix (for instance, the eigenvectors in the basis which the data density matrix diagonalizes, weighted by their eigenvalues).  
Given that mixed states are composed of classical mixtures of pure states, which may not necessarily be orthogonal, it is possible for different sets of pure states to yield the same mixed state. As a consequence, the decomposition of a mixed state into an ensemble of pure states is not unique. Consequently, if only the global mixed state density matrix is provided, individual-level data cannot be recovered.
Moreover, the global objective also offers potential for application in federated learning. In this scenario, the sub-ensembles of each actor may be transformed independently, as their density matrices can be combined additively to generate the full data matrix.

\textbf{Application to genomics studies.}
Combining the specific advantages of our framework, we are particularly driven by potential applications in genomics studies.
Genomics studies involve large-scale datasets that are diverse in terms of data modalities and often contain sensitive information. 
Our framework presents a useful means for addressing key challenges in the integration of quantum computing within such domains by: 1. providing a strategy for the compression of large-scale data into a compact representation, 2.
offering flexible selection of problem-specific objectives for various data types, and 3. providing methods to conceal sensitive training data, for instance in scenarios involving individual-level genomics and clinical data.

\textbf{Relation to previous VAE and QAE models.}
We note that our framework generalizes many aspects of previous VAE and QAE frameworks, such as classical VAE, $\beta$-VAE, Wasserstein VAE, quantum autoencoders, as well as hybrid QVAE models as discussed.  However, we emphasize that our approach is not a straightforward analogue of any specific previous model. Rather, our framework is characterized by the introduction of a general form of training objective, which allows a regularization term to be introduced based on any of the quantum divergences outlined, our principled use of mixed quantum states in the input and latent spaces, and our proposal of a new quantum analogue to the ELBO bound, based on the viewpoint of our framework as a fully quantum generative model of mixed state data.  From our proposed bound and general formulation, we derive the regularization terms used above, as well as quantum analogues to a host of classical models in a uniform way, in addition to the novel class of global loss functions, containing analogues for each of the specific losses above.  The framework is also intended to be extendable in the sense that it provides a principled way of deriving models based on new divergences and combinations of divergences as appropriate to diverse task settings.

\section{Implementation on near-term quantum devices}
It is important to note that implementing our framework on NISQ hardware presents challenges not addressed in this manuscript, such as implementing circuits to input amplitude-encoded state vectors \cite{AmpEncodingCircuit}, reading out latent mixed states with sufficient accuracy, and storage of the resulting density matrices. Specifically for our framework, efficient methods are needed for the divergence calculations between pairs of quantum states in the objective function, and for quantum state tomography, which is required for latent state readout and storage. For the latter, while there are several generally applicable approaches based on matrix-state tomography \cite{MStomography}, neural-network-based tomography \cite{NNtomography, GenMtomography}, or the efficient calculation of density matrix properties \cite{Huang_2020}, the question remains of how well these methods scale for the states learned by our model. 
Frameworks like ours, which use mixed states, may encounter practical difficulties due to the large number of parameters required to fully characterize such states, which in turn impacts the number of state samples needed for accurate readout. In our current simulation-based implementation, the latent states are represented by a small number of qubits, but scaling up could demand the incorporation of additional methods when applied to NISQ hardware. However, if computational efficiency is prioritized, one could constrain the $\zeta$-QVAE model and the choice of $\zeta_\gen$ to handle only pure states, for which procedures like state tomography are known to be less complex. This can be achieved by, for example, projecting the latent mixed state to its most probable eigenstate and using the zero state as $\zeta_\gen$.

Realizing the full benefit of our approach will likely require parallel advancements in state preparation/encoding, storage and tomography. While addressing these common quantum computing challenges is beyond the scope of this manuscript, we do recognize that a full evaluation of the efficiency of our algorithm will necessitate understanding the total computational overhead of all the components acting in concert.

We also recognize that other well-known issues that need to be considered in training datasets, such as the “Barren Plateaus” (BP) effect, would impact our algorithm. In the BP effect, the use of randomly parameterized unitary gates may lead to situations where the average gradient is essentially zero over large swathes of parameter space, preventing the algorithm from finding optima \cite{mcclean2018barrenplateaus}. We note that the BP effect would come into play for our approach when (a) expanding the number of auxiliary qubits (as noted at the end of \autoref{sec:sec5_regularization}) and (b) increasing the number of layers in the encoder and/or decoder circuits. In our current implementations, we kept the number of layers at 3 or lower, but one could imagine cases where the complexity and dimensionality of input datasets requires much deeper circuits. While some strategies to reduce the BP effect have been proposed, such as the use of problem-specific ansatzes (for example, Refs. \cite{a12020034} and \cite{PRXQuantum.1.020319}) , mitigating the BP effect for general circuits is an open problem \cite{Larocca2022diagnosingbarren}. Recent general-purpose proposals include the application of Geometric Quantum Machine Learning (GQML) methods where the known symmetries of the dataset are incorporated into circuit design \cite{ragone2023representation}. Such frameworks could be applied to the design of the encoder and decoder circuits if the dataset has clearly identifiable symmetries. Additionally, we suggest that, given that the primary goal of a quantum compression algorithm is to reduce the dimensionality of the problem for downstream storage and computational efficiency, we foresee that the subsequent compressed states may help reduced the BP effect in downstream circuits by reducing the overall dimensionality of the problem.

\section{Discussion}

We have introduced a novel fully quantum VAE architecture, named $\zeta$-QVAE, which utilizes mixed-state latent representation and provides a flexible framework in which a wide range of quantum reconstruction losses and regularizers can be combined in a unified way.  Further, a theoretical analysis can be given of the objective functions we introduce, which optimize a novel quantum analogue of the ELBO bound underlying the classical VAE. A notable feature of our framework is that mixed states are treated analogously to classical distributions, significantly generalizing previous QAE architectures.  Our results show that our model outperforms classical and alternative QAE models with matched architectures on reconstruction and classification tasks.  

In our experimentation, we demonstrated that the full utilization of the capacity of the latent space, which can be controlled by the parameter $\beta$, is crucial for achieving high-quality latent representations.
We further demonstrated how to fine-tune the trade-off between reconstruction and regularization, and how this allows our model to find the right balance between the two terms to optimize its latent representations for preserving the relational structure of data points and for improved performance in down-stream classification tasks.

We found that there is a complex interplay between regularization and model architecture (including circuit complexity, latent space dimensionality and the inclusion of auxiliary qubits) in determining performance on downstream tasks. Moreover, we have shown the advantage of using general quantum operations between mixed states via auxiliary qubits in our architecture, which increase representational capacity of the model, allowing the dimensionality of the latent and output states to be decoupled.

We further show that our model performs consistently well when trained using a global mixed-state to represent the data, as opposed to individual pure states per data point, thus indicating promising application potential in private and federated learning settings.

With such considerations in mind, we propose that our framework may be ideally suited to constructing quantum models in application areas involving large-scale, heterogeneous and potentially privacy-aware dataset such as genomics. While challenges like data embedding, storage and state tomography remain, our model shows significant practical potential, particularly as advancements in these areas continue to address these obstacles.
In future work, we intend to further investigate how to utilize the observed interaction between model architecture, explicit and implicit regularization, and downstream task performance from the point of view of representational complexity \cite{schuld2021machine}. Further, we intend to investigate how explicit privacy guarantees and federated versions of our approach may be derived for training our model based on our global objective.  Finally, we will investigate the potential of our model to provide efficient compression of intrinsically quantum sources, and implementations of our approach on quantum hardware.

\section*{Code availability}

The code to run $\zeta$-QVAE is available at \texttt{https://github.com/gersteinlab/QVAE.git}. 

\section*{Acknowledgement}
We acknowledge support from the NIH and from the AL Williams Professorship funds. We would also like to thank Huan-hsin Tseng and Aram Harrow for valuable discussions.

\printbibliography

\section{Appendix A}

We provide here further details and proofs regarding the theoretical properties of our framework.  The first relates to the number of qubits required to achieve arbitrary mappings in our encoder and decoder:\\

\noindent \textbf{Proposition 1:}  Setting $N_A = N_B = N_X + 2N_Z$ is sufficient to allow arbitrary pairs of quantum operations $(\mathcal{E},\mathcal{D})$ to be learned in our framework.\\

\noindent \textit{Proof:}  An arbitrary quantum channel $T(.)$ between Hilbert spaces $\mathcal{A}$ and $\mathcal{B}$ may be represented by a unitary transformation $U$ on $\mathcal{A}\otimes\mathcal{B}\otimes\mathcal{C}$:

\begin{eqnarray}\label{prop1}
T(\rho) &=& \Tr_{AC}(U^{-1}(\rho\otimes \ket{\psi_{BC}}\bra{\psi_{BC}})U) 
 \end{eqnarray}

\noindent where $\ket{\psi_{BC}}$ is an arbitrary pure state in $\mathcal{B}\otimes\mathcal{C}$, $\Tr_{AC}$ denotes the trace over $\mathcal{A}\otimes\mathcal{C}$, and $\mathcal{C}$ is an environment with dimension equal to the rank of the Choi matrix representation of $T(.)$, using the Stinespring dilation (see Th. 4.8, \cite{muller2023}).  Since the trace operations in our circuit definitions (Eq. \ref{eq_encDecDef}) are over the final qubits, a final unitary permutation of the qubits may be appended to $U$ in Eq. \ref{prop1}, so that those of $\mathcal{B}$ are mapped to the initial qubits of $\mathcal{A}$ to match the circuit definition in Eq. \ref{eq_encDecDef}.  Since an arbitrary channel between $\mathcal{A}$ and $\mathcal{B}$ may be represented by a Choi matrix of rank between 1 and $\text{dim}(\mathcal{A})\cdot\text{dim}(\mathcal{B})$, $\text{dim}(\mathcal{C})$ is at most $2^{N_X}\cdot 2^{N_Z}$ for input and output spaces $X$ and $Z$ respectively in the encoder (or $Z$ and $X$ in the decoder), and hence may be represented by $\log_2 (2^{N_X+N_Z})=N_X+N_Z$ qubits (in both encoder and decoder).  Hence, $U$ is over a space of dimension $2^{N_X}\cdot 2^{N_Z} \cdot 2^{N_X+N_Z} = 2^{2N_X + 2N_Z}$, and the total number of auxiliary qubits required in both encoder and decoder are $N_A=N_B=(2N_X + 2N_Z)-N_X=N_X + 2N_Z$. \qed \\

\noindent Second, we show that, as in the classical case, the regularized reconstruction loss objective we use is also a lower-bound on the negative quantum relative entropy (the analogue of the classical log-likelihood), when using the quantum relative entropy for both the reconstruction and regularization terms in our objective, and setting $\beta=1$ and $\epsilon=0$.\\

\noindent \textbf{Proposition 2:}  $-S(\rho_\g|\sigma_\gen) \geq -S(\rho_\g|\sigma_\g) - S(\zeta_\g|\zeta_\gen)$\\

\noindent \textit{Proof:} We let $\rho_\g=\sum_i p_i \ket{v_i}\bra{v_i}$, $\zeta_\gen=(1/2^{N_Z})\sum_j \ket{w_j}\bra{w_j}$ and $\zeta_\g=\mathcal{E}(\rho_\g)=\sum_j q_j \ket{w_j}\bra{w_j}$.  Notice that we choose to express $\zeta_\gen$ in the same basis as $\zeta_\g$, which is possible, since the former is the maximally mixed state, which diagonalizes in any basis.  We can express the LHS of the proposition as:

\begin{eqnarray}\label{eq_prop2a}
-S(\rho_\g|\sigma_\gen) &=& \Tr{\rho_\g \log \sigma_\gen} + S(\rho_\g) \nonumber \\
&=& \sum_i p_i \Tr{\ket{v_i}\bra{v_i} \log \sigma_\gen} + S(\rho_\g)
\end{eqnarray} 

\noindent To derive the proposition, we will bound each of the summands $\Tr{\ket{v_i}\bra{v_i} \log \sigma_\gen}$.  We begin by observing the following:

\begin{eqnarray}\label{eq_prop2b}
\Tr{\ket{v_i}\bra{v_i} \sigma_\gen} &=& \mathbb{E}_{j\sim\text{Categ}(1/2^{N_Z})}[\Tr{\ket{v_i}\bra{v_i} \mathcal{D}(\ket{w_j}\bra{w_j})}] \nonumber \\
&=& \mathbb{E}_{j\sim\text{Categ}(q_1...q_{2^{N_Z}}})[\Tr{\ket{v_i}\bra{v_i} \mathcal{D}(\ket{w_j}\bra{w_j})}\cdot\frac{2^{-N_Z}}{q_j}] \nonumber \\
&=& \Tr{\ket{v_i}\bra{v_i}\mathbb{E}_{j\sim\text{Categ}(q_1...q_{2^{N_Z}}})[ \mathcal{D}(\ket{w_j}\bra{w_j})\cdot\frac{2^{-N_Z}}{q_j}]} \nonumber \\
\end{eqnarray} 

\noindent Hence, introducing logs and applying Jensen's trace inequality (lines 2-3), we have:

\begin{eqnarray}\label{eq_prop2c}
&& \Tr{\ket{v_i}\bra{v_i} \log \sigma_\gen} \nonumber \\
&&= \Tr{\ket{v_i}\bra{v_i}\log \mathbb{E}_{j\sim\text{Categ}(q_1...q_{2^{N_Z}}})[ \mathcal{D}(\ket{w_j}\bra{w_j})\cdot\frac{2^{-N_Z}}{q_j}]} \nonumber \\
&&\geq \Tr{\ket{v_i}\bra{v_i} \mathbb{E}_{j\sim\text{Categ}(q_1...q_{2^{N_Z}}})[\log \mathcal{D}(\ket{w_j}\bra{w_j})\cdot\frac{2^{-N_Z}}{q_j}]} \nonumber \\
&&= \Tr{\ket{v_i}\bra{v_i} \mathbb{E}_{j\sim Q}[\log \mathcal{D}(\ket{w_j}\bra{w_j})]} - \mathbb{E}_{j\sim Q}[\log q_j] + \log 2^{-N_Z} \nonumber \\
&&= \Tr{\ket{v_i}\bra{v_i} \log \sigma_\g} + S(\zeta_\g) - S(\zeta_\gen) 
\end{eqnarray} 

\noindent Substituting \autoref{eq_prop2c} into \autoref{eq_prop2a} and summing across $i$, we thus have:

\begin{eqnarray}\label{eq_prop2d}
-S(\rho_\g|\sigma_\gen) &\geq& \sum_i p_i (\Tr{\ket{v_i}\bra{v_i} \log \sigma_\g} + S(\zeta_\g) - S(\zeta_\gen)) + S(\rho_\g) \nonumber \\
&=& -S(\rho_\g|\sigma_\g) + S(\zeta_\g) - S(\zeta_\gen)
\end{eqnarray} 

\noindent and the proposition follows, since $S(\zeta_\g|\zeta_\gen) = S(\zeta_\gen) - S(\zeta_\g)$. \qed \\

\noindent Finally, we show that our global and local objectives are equivalent for linear divergences in the following sense:\\

\noindent \textbf{Proposition 3:}  Our global and local objectives have identical minimizers for $\mathcal{E}$ and $\mathcal{D}$, when they can be expressed in the form given in \autoref{eq_objsAlt}, and $\mathcal{L}'_1$ and $\mathcal{L}_2$ are linear functions their first arguments.\\

\noindent \textit{Proof:} We can express $\rho_\g=(1/N) \sum_i \rho_i$, and $\zeta_\g=(1/N) \sum_i \mathcal{E}(\rho_i) = (1/N) \sum_i \zeta_i$, where $\rho_i$ are the pure states associated with each data-point, and $\zeta_i$ are the associated mixed-state latent representations.  Hence, if $\mathcal{L}'_1$ and $\mathcal{L}_2$ are linear in their first arguments, we have:

\begin{eqnarray}\label{eq_prop3}
\mathcal{L}'_\g(\theta_e,\theta_d,\beta) &=& \mathcal{L}'_1(\rho_\g,\mathcal{E}(\theta_e),\mathcal{D}(\theta_d)) + \beta \mathcal{L}_2(\zeta_\g,\zeta_\gen) \nonumber \\
&=& (1/N) \sum_i \mathcal{L}'_1(\rho_i,\mathcal{E}(\theta_e),\mathcal{D}(\theta_d)) + (1/N) \sum_i \beta \mathcal{L}_2(\zeta_i,\zeta_\gen) \nonumber \\
&=& (1/N) \mathcal{L}'_\inst(\theta_e,\theta_d,\beta)
 \end{eqnarray} 

\noindent Hence, the two objectives are equivalent up to the factor $(1/N)$, leading to identical minimizers. \qed \\

\noindent In particular, Prop. 3 implies that setting $\mathcal{L}'_1$ to the form given in \autoref{eq_altQW} for the Quantum Wasserstein loss, and $\beta=0$ (or setting $\mathcal{L}_2$ to the Quantum Wasserstein loss with respect to the $\zeta_\gen$), results in identical global and local optimization problems.

\end{document}